\documentclass[aps,pre,twocolumn,superscriptaddress]{revtex4-1}
\usepackage{graphicx}
\usepackage{amssymb}
\usepackage{amsmath}
\usepackage{kotex}
\usepackage[mathlines]{lineno}
\usepackage{color}

\usepackage{xcolor}

\newcommand{\skb}[1]{\textcolor{black}{#1}}

\usepackage{multirow}
\usepackage{chngcntr}

\let\oldsubequations\subequations
\let\oldendsubequations\endsubequations

\renewenvironment{subequations}
  {\linenomathNonumbers\oldsubequations}
  {\oldendsubequations\endlinenomath}

\usepackage{tikz-cd}

\begin{document}
\title{Indirect reciprocity as a dynamics for weak balance}
\author{Minwoo Bae}
\affiliation{Research Institute for Basic Sciences, Pukyong National University, Busan 48513, Korea}
\altaffiliation{Current address: Department of Complexity Science and Engineering, The University of Tokyo, Chiba 277-8561, Japan}
\author{Takashi Shimada}
\email[]{shimada@sys.t.u-tokyo.ac.jp}
\affiliation{Department of Systems Innovation, Graduate School of Engineering, The University of Tokyo, 7-3-1 Hongo, Bunkyo-ku, Tokyo 113-8656, Japan}
\author{Seung Ki Baek}
\email[]{seungki@pknu.ac.kr}
\affiliation{Department of Scientific Computing, Pukyong National University, Busan 48513, Korea}

\begin{abstract}
A social network is often divided into many factions. People are friends within each faction, while they are enemies of the other factions, and even my enemy's enemy is not necessarily my friend. This configuration can be described in terms of a weak form of structural balance.
Although weak balance explains a number of real social networks, which dynamical rule achieves it has remained relatively unexplored. In this work, we show that the answer can be found in the field of indirect reciprocity, which assumes that people assess each other's behavior and choose how to behave to others based on the assessment according to a social norm. We begin by showing that weak structural balance is equivalent to stationarity when the rule is given by a norm called `judging'. By analyzing its cluster dynamics of merging, fission, and migration induced by assessment error in complete graphs, we obtain the cluster size distribution in a steady state, which shows the coexistence of a giant cluster and smaller ones.
This study suggests that indirect reciprocity can provide insight into the interplay between a norm that individuals abide by and the macroscopic group structure in society.
\end{abstract}

\maketitle


Judgmental thinking seems to be a universal instinct with which most of us are born. Even infants evaluate each other's behavior~\cite{hamlin2007social}, and their judgment is so broad and conclusive that when they see someone violate moral principles, their inference easily jumps to the wrongdoer's moral character itself~\cite{ting2021toddlers}.
In the field of indirect reciprocity~\cite{nowak1998evolution,ohtsuki2004should,nowak2005evolution,ohtsuki2006leading,hilbe2018indirect,schmid2023quantitative,murase2024computational},
researchers have used a mathematical characterization of judgmental behavior, according to which \skb{society can be governed by a norm called `judging'~\cite{kessinger2023evolution,le2025response}. As the name indicates, it has a high degree of similarity to `stern judging'~\cite{pacheco2006stern,santos2018social,santos2021complexity}, which says that one should not cooperate with the bad, but only the good.
For clarity, we map good (G) and cooperation (C) to $+1$, as well as bad (B) and defection (D) to $-1$, and define $\sigma_{ij}=\pm 1$ as a dynamic variable assigned to every link, say, from player $i$ to $j$, to represent the player $i$'s assessment of $j$.
If $\sigma_{ij}=+1$, the link from $i$ to $j$ is called positive, while $\sigma_{ij}=-1$ means that the link is negative. In the donation game, one player plays the role of a donor, and another player plays the role of a recipient. The donor chooses to cooperate with the recipient or defect, and other players observe the interaction to assess the donor.
Then, judging can be expressed as follows:
\begin{equation}
\sigma_{od}' =
\left\{
\begin{array}{ll}
-1 & \text{if $\sigma_{od} = \sigma_{or}=\sigma_{dr}=-1$}\\
\sigma_{or} \cdot \sigma_{dr} & \text{otherwise},
\end{array}
\right.
\label{eq:judging}
\end{equation}
where $o$, $d$, and $r$ indicate an observer, the donor, and the recipient, respectively, and the prime on the left-hand side means an updated value. According to judging, the donor's action to the recipient should be perfectly correlated with $\sigma_{dr}$, but when the observer assesses the donor, $\sigma_{od}'$ is not determined solely by the donor's action (i.e., $\sigma_{dr}$) but is usually modified by how the observer regards the recipient ($\sigma_{or}$).
Note the only exception \textemdash A bad donor's defection against a bad recipient is again judged as bad, which means that my enemy's enemy is not necessarily my friend~\cite{suppl}.} Thus, it should not be surprising that judging tends to create enemies rather than friends.
This norm of judgment has been regarded as relatively marginal due to its poor performance in promoting cooperation when the assessment is private~\cite{hilbe2018indirect,fujimoto2022reputation,fujimoto2023evolutionary}.
However, a social norm can protect itself from changes, as it makes expectation and action reinforce each other~\cite{mackie2015social}, and this may well be the case even if the norm is not particularly cooperative.
Thus, if we accept it as the {\it status quo} and examine its consequences on macroscopic scales, they could have practical implications, and this is our point of view throughout this work.

\begin{figure}
\includegraphics[width=0.49\columnwidth]{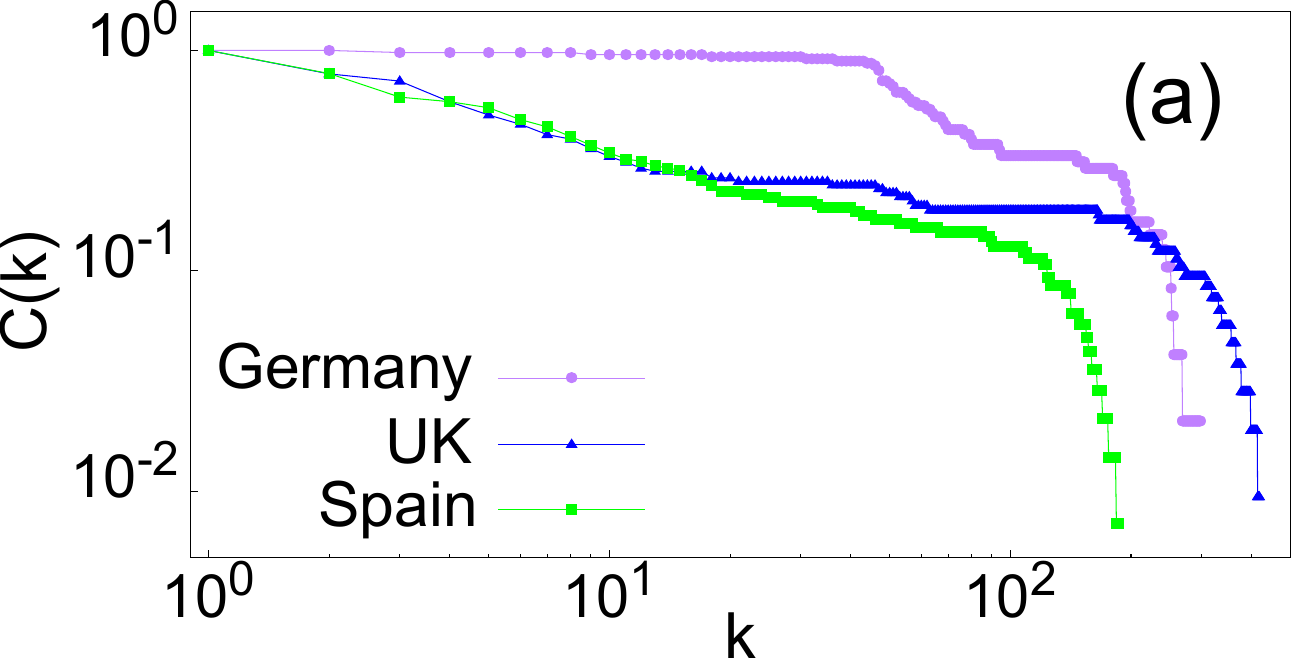}
\includegraphics[width=0.49\columnwidth]{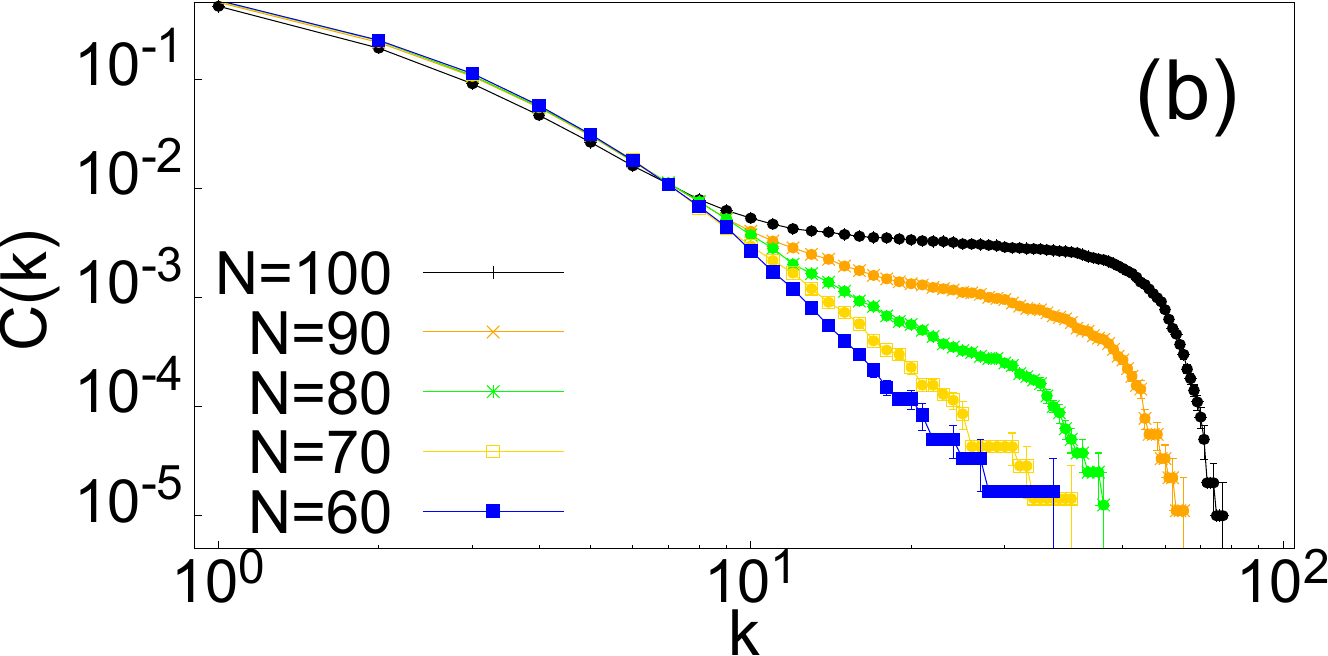}
\caption{(a) Respective cumulative distributions~\cite{codes2} of seats in the parliaments of three countries: Germany (1990 \textendash 2017), the United Kingdom (1983 \textendash 2019), and Spain (1989 \textendash 2020)~\cite{fontan2021signed}.
The plateau up to $k \sim 50$ in the German data may be due to the electoral threshold, which bars small parties from access to the parliament.
(b) Cumulative distribution of cluster sizes in our model
on complete graphs, each with a different number of vertices. We initially start from a random configuration with an equal probability of positive and negative links and let it evolve according to the judging norm until it reaches a weakly balanced configuration. From then on, we attempt transitions among weakly balanced configurations according to the probabilities $P^\ast(m)$, $Q(m)$, $P(m,n)$, and $R(m,n)$ for $2\times 10^9$ times. The system converges to the same distribution regardless of the initial probability of positive links. We have taken the average values and error bars from $10^3$ samples.
}
\label{fig:distribution}
\end{figure}

In the context of social structure, moral judgment plays an ambivalent role. Shared moral values have often been claimed to contribute positively to social cohesion, but the actual effect can be rather complicated~\cite{breidahl2018shared}, and those who conform to a moral norm may even stigmatize those who do not~\cite{tauber2018moralized}.
Politics is one such example closely related to moral judgments, and one of the most common examples of antagonistic group structure in society would be the formation of political parties. 
In fact, empirical studies suggest that political orientations are even more stable than moral intuitions, which implies that our political position might be the true driving force of our moral judgments~\cite{hatemi2019ideology,bakker2021reconsidering}.
In Fig.~\ref{fig:distribution}(a), we show the respective cumulative distributions of seats in the parliaments of Germany, the United Kingdom, and Spain~\cite{fontan2021signed}, which have the largest parliaments among European countries with high human freedom scores~\cite{cato2023human}.
To explain the existence of giant clusters in these broad distributions, one could attempt to construct a phenomenological model of human behavior assuming the probabilities of merging, fission, and migration. However, we would like to propose that it can also be done at a deeper level of social norms, by which one judges another as good or bad.

How does a social norm affect such a group structure?
It is already known that the dynamics of stern judging becomes stationary if and only if Heider's structural balance~\cite{heider1946attitudes} is achieved~\cite{oishi2013group,oishi2021balanced,bae2024exact}. According to the structure theorem~\cite{,harary1953notion,cartwright1956structural}, a balanced configuration consists of two antagonistic groups, within each of which the individuals are positively related. Balance theory can therefore explain, for example, how two allied forces form in the case of warfare~\cite{marvel2011continuous}. However, except for such an extreme conflict, a weak version of structural balance~\cite{easley2010networks} is more favored on social networks~\cite{szell2010multirelational,leskovec2010signed}, and the weak balance is obtained by relaxing the condition that my enemy's enemy is my friend. The corresponding weak version of the structure theorem states that a weakly balanced configuration consists of an arbitrary number of antagonistic groups~\cite{easley2010networks}.
Despite the ubiquity of weak balance, how to achieve it through a dynamical rule remains relatively unexplored, compared to extensive studies on Heider's original balance concept~\cite{malarz2022mean,woloszyn2022thermal,malarz2023thermal,kulakowski2005heider}.


In this work, we will show that judging provides the rule that organizes a weakly balanced configuration. To our knowledge, this is the first report on a dynamical process to achieve weakly balanced configurations as fixed points despite the ubiquity of weak balance in real social networks.
To escape from a weakly balanced configuration, we introduce an assessment error which induces transitions among weakly balanced configurations with well-defined probabilities. By calculating the probabilities, we obtain a coarse-grained description of the judging dynamics at the group level, that is, how groups split, merge, and exchange their members.
The resulting steady-state distribution of group sizes shows a macroscopic consequence of the judging norm and can be compared with group structures in empirical data, such as shown in Fig.~\ref{fig:distribution}.


Consider a complete directed graph of $N$ vertices. Each vertex corresponds to an individual agent, \skb{and the link from a vertex $i$ to another vertex $j$ is given $\sigma_{ij} = \pm 1$ as defined above.} At each time step, we choose a random pair of vertices as a donor and a recipient, respectively. Every individual has the same probability of being a donor, and it is also true for a recipient. The donor and recipient can be the same individual for mathematical convenience, but this probability is negligible when $N$ is large. A weakly balanced configuration is stationary under L8 regardless of self-assessments, so self-assessments are not regarded as relevant degrees of freedom in this work.
All individuals in the population observe the interaction between the donor and the recipient to assess the donor according to the judging norm.
With a small probability $\epsilon$, an observer's assessment of the donor can be flipped from good to bad and vice versa.



\skb{The updating rule in Eq.~\eqref{eq:judging} is equivalent to}
\begin{eqnarray}
\sigma_{ij}' &=& \frac{1}{4} (\sigma_{ij} \sigma_{jk} \sigma_{ik} - \sigma_{ij} \sigma_{jk} - \sigma_{ij} \sigma_{ik})\nonumber\\
&&+ \frac{1}{4} (3 \sigma_{jk} \sigma_{ik} + \sigma_{ij} + \sigma_{jk} + \sigma_{ik} -1),
\label{eq:station}
\end{eqnarray}
when $i$, $j$, and $k$ are the observer, the donor, and the recipient, respectively.
In stationarity, we must have $\sigma_{ij} = \sigma_{ij}'$
for every triad of vertices $i$, $j$, and $k$.
Let us define a detector function for weak balance as follows:
\begin{eqnarray}
W(x,y,z) &\equiv& \frac{1}{4} (1-xyz) (xy + zx + yz - 1)\nonumber\\
&&+ \frac{1}{2} (1+xyz)\\
&=&
\left\{
\begin{array}{ll}
-1 &
\text{if }(x,y,z) \in U,
\\
+1 & \text{otherwise},\nonumber
\end{array}
\right.
\end{eqnarray}
where $U \equiv \{(-1,1,1), (1,-1,1), (1,1,-1)\}$.
Using this detector function, we can easily prove the equivalence between stationarity and weak balance.
That is, if $\sigma_{ij}' = \sigma_{ij}$ everywhere [Eq.~\eqref{eq:station}], it is straightforward to see that $W(\sigma_{ij}, \sigma_{jk}, \sigma_{ik})=+1$, which proves that stationarity implies weak balance. In addition, for each of the five cases where $W(\sigma_{ij}, \sigma_{jk}, \sigma_{ik})=+1$, we find that $\sigma_{ij}' = \sigma_{ij}$, hence the stationarity.

To describe a group structure in mathematical terms, we define a cluster as a maximal clique with respect to positive links. The size of a cluster is equal to the number of vertices inside it.
If only a single cluster exists, it is called `paradise'.
A weakly balanced configuration in a complete graph can be divided into an arbitrary number of clusters in such a way that every pair of two vertices belonging to different clusters is connected by a negative link~\cite{easley2010networks}.
To obtain a basic picture of the cluster dynamics under judging, assume that we have a weakly balanced configuration composed of three clusters as denoted by
$C=\{\{v_1,\ldots,v_{n}\},\{v_{n+1},\ldots,v_{n+m}\},\{v_{n+m+1}, \ldots, v_N\}\}$.
When $v_n$ erroneously regards one of its friends, say $v_1$, as bad, the full enumeration of possible trajectories shows that
the system has only two possibilities: One is to return to the original configuration $C$. It occurs, for example, when $v_n$ sees $v_1$ helping one of its friends from $v_2$ to $v_{n-1}$. The other possibility is to arrive at another weakly balanced configuration $C'=\{\{v_1, \ldots, v_{n-1}\}, \{v_{n}\}, \{v_{n+1}, \ldots, v_{n+m}\}, \{v_{n+m+1}, \ldots, v_N\}\}$,
in which $v_n$ forms a new cluster by itself,
which occurs, for example, when $v_n$ refuses to help $v_1$ and loses reputation from $v_2, \ldots, v_{n-1}$, who in turn refuse to help $v_n$ as a punishment.
If $v_n$ in the configuration $C$ makes a different kind of mistake by judging an enemy, say $v_{n+1}$, as good, the final configuration can be $C$ or $C'$ or $C'' = \{\{v_1, \ldots, v_{n-1}\}, \{v_{n}, v_{n+1}, \ldots, v_{n+m}\}, \{v_{n+m+1}, \ldots, v_N\}\}$, where $v_n$ has migrated to $v_{n+1}$'s cluster.
The trajectory from $C$ to $C''$ is observed, for example, when $v_n$ helping $v_{n+1}$ gains a good reputation from $v_{n+1}, \ldots, v_{n+m}$, who now help $v_n$, while $v_n$'s old friends $v_1, \ldots, v_{n-1}$ refuse to help $v_n$ considering its collaboration with another group.
The process from $C$ to $C'$ will be called fission, and the other process from $C$ to $C''$ will be called migration henceforth.
Note that the last cluster denoted by $\{v_{n+m}, \ldots, v_N\}$ represents all the clusters that are not involved in the mistake committed by $v_n$, and it turns out that they remain bystanders throughout the subsequent process. This implies that we may focus only on the clusters involved with the error during every single process.

Every time the system reaches a weakly balanced configuration through judging, we introduce an assessment error at a random link to let it escape from this absorbing state. Thus, each assessment error defines the unit of time in this dynamics among weakly balanced configurations. More precisely speaking, if $\epsilon$ denotes the probability of assessment error, the time scale $O(1/\epsilon)$ between two consecutive errors is assumed to be much longer than the typical time scale for the system to reach a weakly balanced configuration.
Here we assume that assessment errors occur equally probably at the links for simplicity, but the actual probability has to be estimated to compare our calculations with field observations more accurately.

\begin{figure}
\includegraphics[width=0.9\columnwidth]{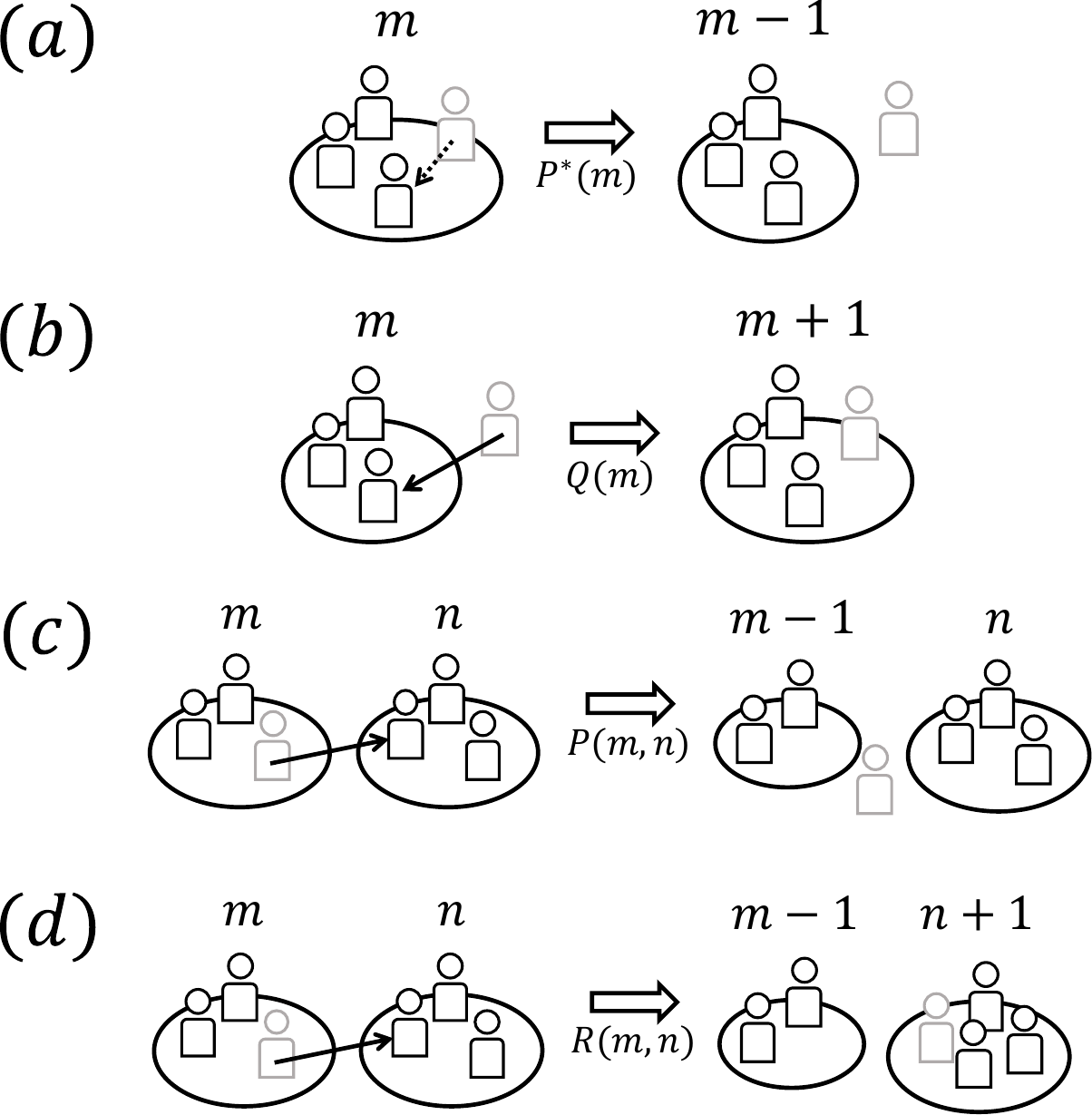}
\caption{\skb{Conditional probabilities for cluster dynamics defined in the main text. Individuals enclosed by a circle means that they belong to the same cluster, and the symbols such as $m$ and $n$ mean the size of each cluster. The dashed arrow is an erroneous bad assessment, and the solid arrows are erroneous good assessment.}
}
\label{fig:eqs}
\end{figure}

\skb{Let $P^\ast(m)$ denote the conditional probability that a vertex in a cluster of size $m$ separates from the others to form a new single-vertex cluster, given that it has committed an error toward a friend in the same cluster, as illustrated in Fig.~\ref{fig:eqs}(a).}
\skb{The inverse process is merging between a single-vertex cluster and another cluster with $m$ vertices, when the single vertex assesses one of its enemies in the other cluster as good by mistake.
Given that the mistake has occurred, the conditional probability of merging is denoted as $Q(m)$, and the process can be depicted as in Fig.~\ref{fig:eqs}(b).}
\skb{To describe the other route of fission, $P(m,n)$ denotes the probability that a vertex in a cluster of size $m$ separates from the others to form a new single-vertex cluster, given that it has committed an error toward an enemy in another cluster of size $n$. The process occurs as depicted in Fig.~\ref{fig:eqs}(c)}.
\skb{The same kind of error may also lead to the migration of the error-committing vertex from the original cluster of size $m$ to the other cluster of size $n$ with probability $R(m,n)$ as shown in Fig.~\ref{fig:eqs}(d).}
%
\begin{figure}
\includegraphics[width=0.49\columnwidth]{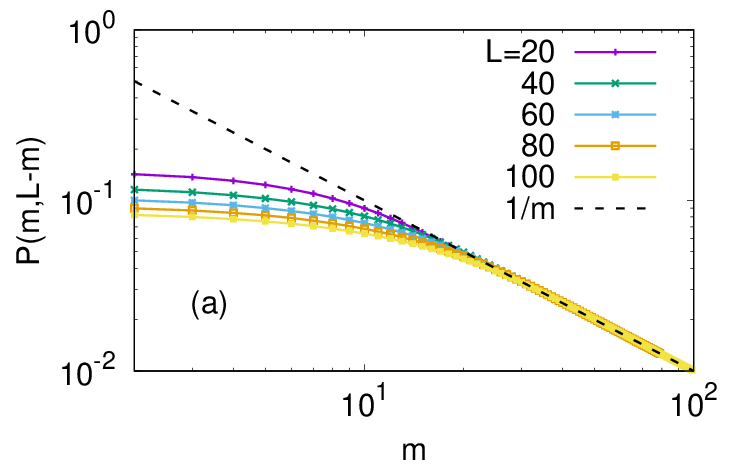}
\includegraphics[width=0.49\columnwidth]{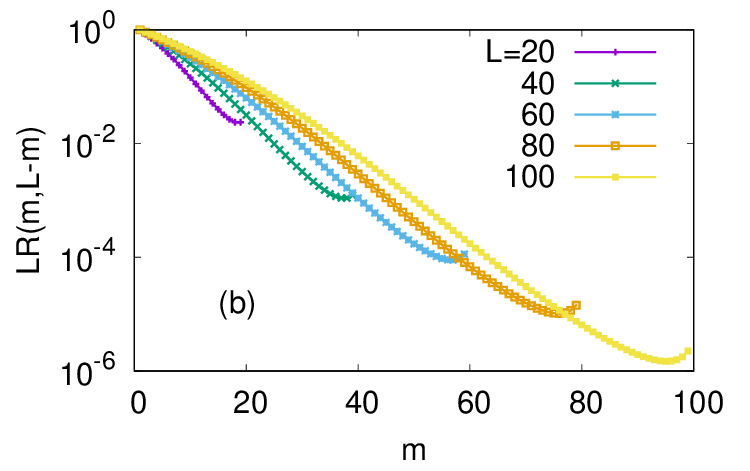}
\caption{(a) Conditional probability~\cite{codes2} of fission given an assessment error [Fig.~\ref{fig:eqs}(c)], when $L \equiv m+n$ is fixed. We have depicted $P^\ast(m) = 1/m$ as a dashed line for comparison. (b) Conditional probability of migration given an assessment error [Fig.~\ref{fig:eqs}(d)], multiplied by $L$ to incorporate $R(1,L-1) \equiv Q(L-1) = 1/L$ as an end point.}
\label{fig:prob}
\end{figure}
We have proved $P^\ast(m)=Q(m-1)=1/m$~\cite{suppl} and developed a numerically exact method to calculate $P(m,n)$ and $R(m,n)$~\cite{suppl,codes2}. Figure~\ref{fig:prob} shows the results when $L\equiv m+n$ is fixed. Note that we have identified $R(1,L-1)$ with $Q(L-1) = 1/L$ because migration is effectively identical to merging if a single-vertex cluster is absorbed into another cluster.
It is also worth noting that $P(m,L-m) \approx P^\ast(m) = 1/m$ when $m \gtrsim O(10)$.


Using the probabilities of fission, merging, and migration obtained above, we calculate the cumulative distribution of cluster sizes, \skb{$C(k) \equiv \int_k^\infty \rho(k') dk'$}, in steady state [Fig.~\ref{fig:distribution}(b)]. Here, the \skb{number density of} a cluster of size $k$ is denoted by \skb{$\rho(k)$}, and the normalization condition is given by \skb{$\sum_k k\rho(k) = 1$}. In the language of percolation, the distribution suggests that the system is in a supercritical phase, where we find a giant cluster that occupies a finite fraction of the system.
This analogy with percolation predicts that the overall frequency of good assessments will be low, although greater than zero, because we have positive links within a finite fraction of the system. This prediction is indeed consistent with a recent study~\cite{fujimoto2024leader}, in which the average frequency is found to be around $30\%$ under judging in the presence of assessment error. This is even lower than that of stern judging, according to which every player can expect good assessments from its friends comprising $50\%$ of the population~\cite{bae2024exact}.

To elucidate the above result, assume that we have a single giant cluster of size $K \gg 1$, \skb{which will be counted separately from the other smaller clusters}. If the number of clusters of size $k$ is denoted by $n_k$, we have
\begin{equation}
K + \sum_k k n_k = N.
\label{eq:total}
\end{equation}
In a steady state, the increase of $n_1$ due to the breakage of the giant cluster is written as follows:
\begin{equation}
\Delta n_1^G = \frac{K^2}{N^2} P^\ast(K) + \sum_{k=1} \frac{(kn_k) K}{N^2} P(K,k),
\label{eq:n1G}
\end{equation}
where the first term comes from an error inside the giant cluster, and the second term comes from an error from a member of the giant cluster toward someone else in another cluster of size $k$.
If we note that $P(K,k) \approx P^\ast(K) =1/K$, it simplifies to $\Delta n_1^G \approx 1/N$, which means that clusters consisting of a single vertex are generated from the giant cluster at a constant rate. When other finite clusters of size $k>1$ break, the contribution can be expressed by
\begin{eqnarray}
\Delta n_1^F &=& \sum_{k=2} n_k \frac{k^2}{N^2} P^\ast(k) \left( 1+\delta_{k,2}\right)\\
&&+ \sum_{k=2} \sum_{k'=1} \frac{(kn_k)(k'n_{k'})}{N^2} P(k,k') \left( 1+\delta_{k,2}\right),\nonumber
\label{eq:n1F}
\end{eqnarray}
where the Kronecker delta takes into account the fact that $n_1$ increases by two when a cluster of $k=2$ breaks. The summation over $k'$ includes the case of $k'=K$. The loss terms of $n_1$ can be written as
\begin{eqnarray}
\Delta n_1^- &=&  \sum_{k=1} \frac{n_1(kn_k)}{N^2} R(1,k) \left( 1+\delta_{k,1}\right)\nonumber\\
&&+ \sum_{k=3} \frac{n_1(kn_k)}{N^2} R(k,1) ,
\label{eq:n1-}
\end{eqnarray}
where the Kronecker delta again expresses the fact that $n_1$ decreases by two when two clusters of $k=1$ merge. As above, the summations over $k$ include the case of $k=K$. In a steady state, $\Delta n_1^G + \Delta n_1^F$ must equal $\Delta n_1^-$.
The change of $n_k$ with $k>1$ can be given in a similar way~\cite{suppl}.
If we neglect all finite clusters of $k>1$, we have
$N + n_1 \approx n_1 + n_1^2$,
which is solved by $n_1 = \sqrt{N} \approx N-K$.
It means that the creation of small clusters from the giant one must be balanced with the reverse process through which smaller clusters are absorbed into the giant one, in addition to the migration of individuals between small clusters. The resulting behavior of $K \propto N$ is consistent with our initial assumption that a giant cluster emerges.


Before concluding, we add that judging is not the only mechanism to achieve a weakly balanced configuration.
Stationarity is equivalent to weak balance in another social norm called `staying' (also known as L7). It is different from judging (L8) only by $\alpha_{GCB}=G$~\skb{\cite{suppl}}. Considering the same difference between L4 and L6 (stern judging), we can say that L7 (staying) is for L8 (judging) what L4 is for L6 (stern judging). In fact, under L7 (staying), the system arrives at paradise in a way similar to L4~\cite{bae2024exact}.
This suggests how a small change in a social norm can induce macroscopic changes throughout the social network.
Weak balance can sometimes be achieved even without a social norm. \skb{Suppose that each individual has to choose a color of clothing from a few given possibilities. Provided that they are friends if and only if they share the same color but are enemies otherwise, such homophily will induce a weakly balanced social network because the network will be split into as many clusters as the number of colors.
However, in the absence of such discrete properties,} we need a specific network dynamics that suppresses unbalanced triangles. One such example is the social inheritance model~\cite{ilany2013structural,ilany2016social}, which has a tendency to promote local clustering.
Compared to our model, a fundamental difference of the social inheritance model is that it does not require equivalence between stationarity and balance, although we have confirmed that it achieves a weak balance in the long run (not shown). As a consequence, the system may be stationary without balance or may continue to change in a balanced configuration.
When weak balance is observed, one could tell its mechanism by referring to the different predictions from those competing explanations, that is, homophily, social inheritance, and the judging norm.
Among them, our norm-based explanation is the one that provides probabilities for cluster dynamics in a numerically exact manner, and hence is open to further scrutiny.

\begin{acknowledgments}
M.B. and S.K.B. acknowledge support by Basic Science Research Program through the
National Research Foundation of Korea (NRF) funded by the Ministry of Education
(NRF-2020R1I1A2071670).
We appreciate the APCTP for its hospitality during the completion of this work.
\end{acknowledgments}


\begin{thebibliography}{43}%
\makeatletter
\providecommand \@ifxundefined [1]{%
 \@ifx{#1\undefined}
}%
\providecommand \@ifnum [1]{%
 \ifnum #1\expandafter \@firstoftwo
 \else \expandafter \@secondoftwo
 \fi
}%
\providecommand \@ifx [1]{%
 \ifx #1\expandafter \@firstoftwo
 \else \expandafter \@secondoftwo
 \fi
}%
\providecommand \natexlab [1]{#1}%
\providecommand \enquote  [1]{``#1''}%
\providecommand \bibnamefont  [1]{#1}%
\providecommand \bibfnamefont [1]{#1}%
\providecommand \citenamefont [1]{#1}%
\providecommand \href@noop [0]{\@secondoftwo}%
\providecommand \href [0]{\begingroup \@sanitize@url \@href}%
\providecommand \@href[1]{\@@startlink{#1}\@@href}%
\providecommand \@@href[1]{\endgroup#1\@@endlink}%
\providecommand \@sanitize@url [0]{\catcode `\\12\catcode `\$12\catcode
  `\&12\catcode `\#12\catcode `\^12\catcode `\_12\catcode `\%12\relax}%
\providecommand \@@startlink[1]{}%
\providecommand \@@endlink[0]{}%
\providecommand \url  [0]{\begingroup\@sanitize@url \@url }%
\providecommand \@url [1]{\endgroup\@href {#1}{\urlprefix }}%
\providecommand \urlprefix  [0]{URL }%
\providecommand \Eprint [0]{\href }%
\providecommand \doibase [0]{http://dx.doi.org/}%
\providecommand \selectlanguage [0]{\@gobble}%
\providecommand \bibinfo  [0]{\@secondoftwo}%
\providecommand \bibfield  [0]{\@secondoftwo}%
\providecommand \translation [1]{[#1]}%
\providecommand \BibitemOpen [0]{}%
\providecommand \bibitemStop [0]{}%
\providecommand \bibitemNoStop [0]{.\EOS\space}%
\providecommand \EOS [0]{\spacefactor3000\relax}%
\providecommand \BibitemShut  [1]{\csname bibitem#1\endcsname}%
\let\auto@bib@innerbib\@empty
\bibitem [{\citenamefont {Hamlin}\ \emph {et~al.}(2007)\citenamefont {Hamlin},
  \citenamefont {Wynn},\ and\ \citenamefont {Bloom}}]{hamlin2007social}%
  \BibitemOpen
  \bibfield  {author} {\bibinfo {author} {\bibfnamefont {J.~K.}\ \bibnamefont
  {Hamlin}}, \bibinfo {author} {\bibfnamefont {K.}~\bibnamefont {Wynn}}, \ and\
  \bibinfo {author} {\bibfnamefont {P.}~\bibnamefont {Bloom}},\ }\href@noop {}
  {\bibfield  {journal} {\bibinfo  {journal} {Nature}\ }\textbf {\bibinfo
  {volume} {450}},\ \bibinfo {pages} {557} (\bibinfo {year}
  {2007})}\BibitemShut {NoStop}%
\bibitem [{\citenamefont {Ting}\ and\ \citenamefont
  {Baillargeon}(2021)}]{ting2021toddlers}%
  \BibitemOpen
  \bibfield  {author} {\bibinfo {author} {\bibfnamefont {F.}~\bibnamefont
  {Ting}}\ and\ \bibinfo {author} {\bibfnamefont {R.}~\bibnamefont
  {Baillargeon}},\ }\href@noop {} {\bibfield  {journal} {\bibinfo  {journal}
  {Proc. Natl. Acad. Sci. USA}\ }\textbf {\bibinfo {volume} {118}},\ \bibinfo
  {pages} {e2109045118} (\bibinfo {year} {2021})}\BibitemShut {NoStop}%
\bibitem [{\citenamefont {Nowak}\ and\ \citenamefont
  {Sigmund}(1998)}]{nowak1998evolution}%
  \BibitemOpen
  \bibfield  {author} {\bibinfo {author} {\bibfnamefont {M.~A.}\ \bibnamefont
  {Nowak}}\ and\ \bibinfo {author} {\bibfnamefont {K.}~\bibnamefont
  {Sigmund}},\ }\href@noop {} {\bibfield  {journal} {\bibinfo  {journal}
  {Nature}\ }\textbf {\bibinfo {volume} {393}},\ \bibinfo {pages} {573}
  (\bibinfo {year} {1998})}\BibitemShut {NoStop}%
\bibitem [{\citenamefont {Ohtsuki}\ and\ \citenamefont
  {Iwasa}(2004)}]{ohtsuki2004should}%
  \BibitemOpen
  \bibfield  {author} {\bibinfo {author} {\bibfnamefont {H.}~\bibnamefont
  {Ohtsuki}}\ and\ \bibinfo {author} {\bibfnamefont {Y.}~\bibnamefont
  {Iwasa}},\ }\href@noop {} {\bibfield  {journal} {\bibinfo  {journal} {J.
  Theor. Biol.}\ }\textbf {\bibinfo {volume} {231}},\ \bibinfo {pages} {107}
  (\bibinfo {year} {2004})}\BibitemShut {NoStop}%
\bibitem [{\citenamefont {Nowak}\ and\ \citenamefont
  {Sigmund}(2005)}]{nowak2005evolution}%
  \BibitemOpen
  \bibfield  {author} {\bibinfo {author} {\bibfnamefont {M.~A.}\ \bibnamefont
  {Nowak}}\ and\ \bibinfo {author} {\bibfnamefont {K.}~\bibnamefont
  {Sigmund}},\ }\href@noop {} {\bibfield  {journal} {\bibinfo  {journal}
  {Nature}\ }\textbf {\bibinfo {volume} {437}},\ \bibinfo {pages} {1291}
  (\bibinfo {year} {2005})}\BibitemShut {NoStop}%
\bibitem [{\citenamefont {Ohtsuki}\ and\ \citenamefont
  {Iwasa}(2006)}]{ohtsuki2006leading}%
  \BibitemOpen
  \bibfield  {author} {\bibinfo {author} {\bibfnamefont {H.}~\bibnamefont
  {Ohtsuki}}\ and\ \bibinfo {author} {\bibfnamefont {Y.}~\bibnamefont
  {Iwasa}},\ }\href@noop {} {\bibfield  {journal} {\bibinfo  {journal} {J.
  Theor. Biol.}\ }\textbf {\bibinfo {volume} {239}},\ \bibinfo {pages} {435}
  (\bibinfo {year} {2006})}\BibitemShut {NoStop}%
\bibitem [{\citenamefont {Hilbe}\ \emph {et~al.}(2018)\citenamefont {Hilbe},
  \citenamefont {Schmid}, \citenamefont {Tkadlec}, \citenamefont {Chatterjee},\
  and\ \citenamefont {Nowak}}]{hilbe2018indirect}%
  \BibitemOpen
  \bibfield  {author} {\bibinfo {author} {\bibfnamefont {C.}~\bibnamefont
  {Hilbe}}, \bibinfo {author} {\bibfnamefont {L.}~\bibnamefont {Schmid}},
  \bibinfo {author} {\bibfnamefont {J.}~\bibnamefont {Tkadlec}}, \bibinfo
  {author} {\bibfnamefont {K.}~\bibnamefont {Chatterjee}}, \ and\ \bibinfo
  {author} {\bibfnamefont {M.~A.}\ \bibnamefont {Nowak}},\ }\href@noop {}
  {\bibfield  {journal} {\bibinfo  {journal} {Proc. Natl. Acad. Sci. USA}\
  }\textbf {\bibinfo {volume} {115}},\ \bibinfo {pages} {12241} (\bibinfo
  {year} {2018})}\BibitemShut {NoStop}%
\bibitem [{\citenamefont {Schmid}\ \emph {et~al.}(2023)\citenamefont {Schmid},
  \citenamefont {Ekbatani}, \citenamefont {Hilbe},\ and\ \citenamefont
  {Chatterjee}}]{schmid2023quantitative}%
  \BibitemOpen
  \bibfield  {author} {\bibinfo {author} {\bibfnamefont {L.}~\bibnamefont
  {Schmid}}, \bibinfo {author} {\bibfnamefont {F.}~\bibnamefont {Ekbatani}},
  \bibinfo {author} {\bibfnamefont {C.}~\bibnamefont {Hilbe}}, \ and\ \bibinfo
  {author} {\bibfnamefont {K.}~\bibnamefont {Chatterjee}},\ }\href@noop {}
  {\bibfield  {journal} {\bibinfo  {journal} {Nat. Commun.}\ }\textbf {\bibinfo
  {volume} {14}},\ \bibinfo {pages} {2086} (\bibinfo {year}
  {2023})}\BibitemShut {NoStop}%
\bibitem [{\citenamefont {Murase}\ and\ \citenamefont
  {Hilbe}(2024)}]{murase2024computational}%
  \BibitemOpen
  \bibfield  {author} {\bibinfo {author} {\bibfnamefont {Y.}~\bibnamefont
  {Murase}}\ and\ \bibinfo {author} {\bibfnamefont {C.}~\bibnamefont {Hilbe}},\
  }\href@noop {} {\bibfield  {journal} {\bibinfo  {journal} {Proc. Natl. Acad.
  Sci. USA}\ }\textbf {\bibinfo {volume} {121}},\ \bibinfo {pages}
  {e2406885121} (\bibinfo {year} {2024})}\BibitemShut {NoStop}%
\bibitem [{\citenamefont {Kessinger}\ \emph {et~al.}(2023)\citenamefont
  {Kessinger}, \citenamefont {Tarnita},\ and\ \citenamefont
  {Plotkin}}]{kessinger2023evolution}%
  \BibitemOpen
  \bibfield  {author} {\bibinfo {author} {\bibfnamefont {T.~A.}\ \bibnamefont
  {Kessinger}}, \bibinfo {author} {\bibfnamefont {C.~E.}\ \bibnamefont
  {Tarnita}}, \ and\ \bibinfo {author} {\bibfnamefont {J.~B.}\ \bibnamefont
  {Plotkin}},\ }\href@noop {} {\bibfield  {journal} {\bibinfo  {journal} {Proc.
  Natl. Acad. Sci. USA}\ }\textbf {\bibinfo {volume} {120}},\ \bibinfo {pages}
  {e2219480120} (\bibinfo {year} {2023})}\BibitemShut {NoStop}%
\bibitem [{\citenamefont {Le}\ and\ \citenamefont
  {Baek}(2025)}]{le2025response}%
  \BibitemOpen
  \bibfield  {author} {\bibinfo {author} {\bibfnamefont {Q.~A.}\ \bibnamefont
  {Le}}\ and\ \bibinfo {author} {\bibfnamefont {S.~K.}\ \bibnamefont {Baek}},\
  }\href@noop {} {\bibfield  {journal} {\bibinfo  {journal} {J. Theor. Biol.}\
  }\textbf {\bibinfo {volume} {612}},\ \bibinfo {pages} {112199} (\bibinfo
  {year} {2025})}\BibitemShut {NoStop}%
\bibitem [{\citenamefont {Pacheco}\ \emph {et~al.}(2006)\citenamefont
  {Pacheco}, \citenamefont {Santos},\ and\ \citenamefont
  {Chalub}}]{pacheco2006stern}%
  \BibitemOpen
  \bibfield  {author} {\bibinfo {author} {\bibfnamefont {J.~M.}\ \bibnamefont
  {Pacheco}}, \bibinfo {author} {\bibfnamefont {F.~C.}\ \bibnamefont {Santos}},
  \ and\ \bibinfo {author} {\bibfnamefont {F.~A.~C.}\ \bibnamefont {Chalub}},\
  }\href@noop {} {\bibfield  {journal} {\bibinfo  {journal} {PLoS Comput.
  Biol.}\ }\textbf {\bibinfo {volume} {2}},\ \bibinfo {pages} {e178} (\bibinfo
  {year} {2006})}\BibitemShut {NoStop}%
\bibitem [{\citenamefont {Santos}\ \emph {et~al.}(2018)\citenamefont {Santos},
  \citenamefont {Santos},\ and\ \citenamefont {Pacheco}}]{santos2018social}%
  \BibitemOpen
  \bibfield  {author} {\bibinfo {author} {\bibfnamefont {F.~P.}\ \bibnamefont
  {Santos}}, \bibinfo {author} {\bibfnamefont {F.~C.}\ \bibnamefont {Santos}},
  \ and\ \bibinfo {author} {\bibfnamefont {J.~M.}\ \bibnamefont {Pacheco}},\
  }\href@noop {} {\bibfield  {journal} {\bibinfo  {journal} {Nature}\ }\textbf
  {\bibinfo {volume} {555}},\ \bibinfo {pages} {242} (\bibinfo {year}
  {2018})}\BibitemShut {NoStop}%
\bibitem [{\citenamefont {Santos}\ \emph {et~al.}(2021)\citenamefont {Santos},
  \citenamefont {Pacheco},\ and\ \citenamefont
  {Santos}}]{santos2021complexity}%
  \BibitemOpen
  \bibfield  {author} {\bibinfo {author} {\bibfnamefont {F.~P.}\ \bibnamefont
  {Santos}}, \bibinfo {author} {\bibfnamefont {J.~M.}\ \bibnamefont {Pacheco}},
  \ and\ \bibinfo {author} {\bibfnamefont {F.~C.}\ \bibnamefont {Santos}},\
  }\href@noop {} {\bibfield  {journal} {\bibinfo  {journal} {Philos. Trans. R.
  Soc. B}\ }\textbf {\bibinfo {volume} {376}},\ \bibinfo {pages} {20200291}
  (\bibinfo {year} {2021})}\BibitemShut {NoStop}%
\bibitem [{sup()}]{suppl}%
  \BibitemOpen
  \href@noop {} {}\bibinfo {note} {See Supplemental Material for the
  characterization of the leading eight, the calculation of transition
  probabilities, and the estimate of the giant cluster size.}\BibitemShut
  {Stop}%
\bibitem [{\citenamefont {Fujimoto}\ and\ \citenamefont
  {Ohtsuki}(2022)}]{fujimoto2022reputation}%
  \BibitemOpen
  \bibfield  {author} {\bibinfo {author} {\bibfnamefont {Y.}~\bibnamefont
  {Fujimoto}}\ and\ \bibinfo {author} {\bibfnamefont {H.}~\bibnamefont
  {Ohtsuki}},\ }\href@noop {} {\bibfield  {journal} {\bibinfo  {journal} {Sci.
  Rep.}\ }\textbf {\bibinfo {volume} {12}},\ \bibinfo {pages} {10500} (\bibinfo
  {year} {2022})}\BibitemShut {NoStop}%
\bibitem [{\citenamefont {Fujimoto}\ and\ \citenamefont
  {Ohtsuki}(2023)}]{fujimoto2023evolutionary}%
  \BibitemOpen
  \bibfield  {author} {\bibinfo {author} {\bibfnamefont {Y.}~\bibnamefont
  {Fujimoto}}\ and\ \bibinfo {author} {\bibfnamefont {H.}~\bibnamefont
  {Ohtsuki}},\ }\href@noop {} {\bibfield  {journal} {\bibinfo  {journal} {Proc.
  Natl. Acad. Sci. USA}\ }\textbf {\bibinfo {volume} {120}},\ \bibinfo {pages}
  {e2300544120} (\bibinfo {year} {2023})}\BibitemShut {NoStop}%
\bibitem [{\citenamefont {Mackie}\ \emph {et~al.}(2015)\citenamefont {Mackie},
  \citenamefont {Moneti}, \citenamefont {Shakya},\ and\ \citenamefont
  {Denny}}]{mackie2015social}%
  \BibitemOpen
  \bibfield  {author} {\bibinfo {author} {\bibfnamefont {G.}~\bibnamefont
  {Mackie}}, \bibinfo {author} {\bibfnamefont {F.}~\bibnamefont {Moneti}},
  \bibinfo {author} {\bibfnamefont {H.}~\bibnamefont {Shakya}}, \ and\ \bibinfo
  {author} {\bibfnamefont {E.}~\bibnamefont {Denny}},\ }\href@noop {} {\emph
  {\bibinfo {title} {What are social norms? How are they measured}}},\ \bibinfo
  {type} {Working Paper}\ (\bibinfo  {institution} {University of California
  San Diego Center on Global Justice},\ \bibinfo {address} {San Diego, CA},\
  \bibinfo {year} {2015})\BibitemShut {NoStop}%
\bibitem [{\citenamefont {Breidahl}\ \emph {et~al.}(2018)\citenamefont
  {Breidahl}, \citenamefont {Holtug},\ and\ \citenamefont
  {Kongsh{\o}j}}]{breidahl2018shared}%
  \BibitemOpen
  \bibfield  {author} {\bibinfo {author} {\bibfnamefont {K.~N.}\ \bibnamefont
  {Breidahl}}, \bibinfo {author} {\bibfnamefont {N.}~\bibnamefont {Holtug}}, \
  and\ \bibinfo {author} {\bibfnamefont {K.}~\bibnamefont {Kongsh{\o}j}},\
  }\href@noop {} {\bibfield  {journal} {\bibinfo  {journal} {Eur. Political
  Sci. Rev.}\ }\textbf {\bibinfo {volume} {10}},\ \bibinfo {pages} {97}
  (\bibinfo {year} {2018})}\BibitemShut {NoStop}%
\bibitem [{\citenamefont {T{\"a}uber}(2018)}]{tauber2018moralized}%
  \BibitemOpen
  \bibfield  {author} {\bibinfo {author} {\bibfnamefont {S.}~\bibnamefont
  {T{\"a}uber}},\ }\href@noop {} {\bibfield  {journal} {\bibinfo  {journal}
  {Front. Psychol.}\ }\textbf {\bibinfo {volume} {9}},\ \bibinfo {pages} {909}
  (\bibinfo {year} {2018})}\BibitemShut {NoStop}%
\bibitem [{\citenamefont {Hatemi}\ \emph {et~al.}(2019)\citenamefont {Hatemi},
  \citenamefont {Crabtree},\ and\ \citenamefont {Smith}}]{hatemi2019ideology}%
  \BibitemOpen
  \bibfield  {author} {\bibinfo {author} {\bibfnamefont {P.~K.}\ \bibnamefont
  {Hatemi}}, \bibinfo {author} {\bibfnamefont {C.}~\bibnamefont {Crabtree}}, \
  and\ \bibinfo {author} {\bibfnamefont {K.~B.}\ \bibnamefont {Smith}},\
  }\href@noop {} {\bibfield  {journal} {\bibinfo  {journal} {Am. J. Political
  Sci.}\ }\textbf {\bibinfo {volume} {63}},\ \bibinfo {pages} {788} (\bibinfo
  {year} {2019})}\BibitemShut {NoStop}%
\bibitem [{\citenamefont {Bakker}\ \emph {et~al.}(2021)\citenamefont {Bakker},
  \citenamefont {Lelkes},\ and\ \citenamefont
  {Malka}}]{bakker2021reconsidering}%
  \BibitemOpen
  \bibfield  {author} {\bibinfo {author} {\bibfnamefont {B.~N.}\ \bibnamefont
  {Bakker}}, \bibinfo {author} {\bibfnamefont {Y.}~\bibnamefont {Lelkes}}, \
  and\ \bibinfo {author} {\bibfnamefont {A.}~\bibnamefont {Malka}},\
  }\href@noop {} {\bibfield  {journal} {\bibinfo  {journal} {Am. Political Sci.
  Rev.}\ }\textbf {\bibinfo {volume} {115}},\ \bibinfo {pages} {1482} (\bibinfo
  {year} {2021})}\BibitemShut {NoStop}%
\bibitem [{\citenamefont {Fontan}\ and\ \citenamefont
  {Altafini}(2021)}]{fontan2021signed}%
  \BibitemOpen
  \bibfield  {author} {\bibinfo {author} {\bibfnamefont {A.}~\bibnamefont
  {Fontan}}\ and\ \bibinfo {author} {\bibfnamefont {C.}~\bibnamefont
  {Altafini}},\ }\href@noop {} {\bibfield  {journal} {\bibinfo  {journal} {Sci.
  Rep.}\ }\textbf {\bibinfo {volume} {11}},\ \bibinfo {pages} {5134} (\bibinfo
  {year} {2021})}\BibitemShut {NoStop}%
\bibitem [{\citenamefont {{Cato Institute}}()}]{cato2023human}%
  \BibitemOpen
  \bibfield  {author} {\bibinfo {author} {\bibnamefont {{Cato Institute}}},\
  }\href@noop {} {}\bibinfo {howpublished} {available at
  https://www.cato.org/human-freedom-index/2023 (accessed 2024 Dec
  30)}\BibitemShut {NoStop}%
\bibitem [{cod()}]{codes2}%
  \BibitemOpen
  \href@noop {} {}\bibinfo {howpublished} {Codes to obtain the distributions in
  Fig. 1 and the probabilities in Fig. 3 are available at
  https://github.com/BOS-Bae/Fragmented-Complete-Network.}\BibitemShut {Stop}%
\bibitem [{\citenamefont {Heider}(1946)}]{heider1946attitudes}%
  \BibitemOpen
  \bibfield  {author} {\bibinfo {author} {\bibfnamefont {F.}~\bibnamefont
  {Heider}},\ }\href@noop {} {\bibfield  {journal} {\bibinfo  {journal} {J.
  Psychol.}\ }\textbf {\bibinfo {volume} {21}},\ \bibinfo {pages} {107}
  (\bibinfo {year} {1946})}\BibitemShut {NoStop}%
\bibitem [{\citenamefont {Oishi}\ \emph {et~al.}(2013)\citenamefont {Oishi},
  \citenamefont {Shimada},\ and\ \citenamefont {Ito}}]{oishi2013group}%
  \BibitemOpen
  \bibfield  {author} {\bibinfo {author} {\bibfnamefont {K.}~\bibnamefont
  {Oishi}}, \bibinfo {author} {\bibfnamefont {T.}~\bibnamefont {Shimada}}, \
  and\ \bibinfo {author} {\bibfnamefont {N.}~\bibnamefont {Ito}},\ }\href@noop
  {} {\bibfield  {journal} {\bibinfo  {journal} {Phys. Rev. E}\ }\textbf
  {\bibinfo {volume} {87}},\ \bibinfo {pages} {030801(R)} (\bibinfo {year}
  {2013})}\BibitemShut {NoStop}%
\bibitem [{\citenamefont {Oishi}\ \emph {et~al.}(2021)\citenamefont {Oishi},
  \citenamefont {Miyano}, \citenamefont {Kaski},\ and\ \citenamefont
  {Shimada}}]{oishi2021balanced}%
  \BibitemOpen
  \bibfield  {author} {\bibinfo {author} {\bibfnamefont {K.}~\bibnamefont
  {Oishi}}, \bibinfo {author} {\bibfnamefont {S.}~\bibnamefont {Miyano}},
  \bibinfo {author} {\bibfnamefont {K.}~\bibnamefont {Kaski}}, \ and\ \bibinfo
  {author} {\bibfnamefont {T.}~\bibnamefont {Shimada}},\ }\href@noop {}
  {\bibfield  {journal} {\bibinfo  {journal} {Phys. Rev. E}\ }\textbf {\bibinfo
  {volume} {104}},\ \bibinfo {pages} {024310} (\bibinfo {year}
  {2021})}\BibitemShut {NoStop}%
\bibitem [{\citenamefont {Bae}\ \emph {et~al.}(2024)\citenamefont {Bae},
  \citenamefont {Shimada},\ and\ \citenamefont {Baek}}]{bae2024exact}%
  \BibitemOpen
  \bibfield  {author} {\bibinfo {author} {\bibfnamefont {M.}~\bibnamefont
  {Bae}}, \bibinfo {author} {\bibfnamefont {T.}~\bibnamefont {Shimada}}, \ and\
  \bibinfo {author} {\bibfnamefont {S.~K.}\ \bibnamefont {Baek}},\ }\href@noop
  {} {\bibfield  {journal} {\bibinfo  {journal} {Phys. Rev. E}\ }\textbf
  {\bibinfo {volume} {110}},\ \bibinfo {pages} {L052301} (\bibinfo {year}
  {2024})}\BibitemShut {NoStop}%
\bibitem [{\citenamefont {Harary}(1953)}]{harary1953notion}%
  \BibitemOpen
  \bibfield  {author} {\bibinfo {author} {\bibfnamefont {F.}~\bibnamefont
  {Harary}},\ }\href@noop {} {\bibfield  {journal} {\bibinfo  {journal} {Mich.
  Math. J.}\ }\textbf {\bibinfo {volume} {2}},\ \bibinfo {pages} {143}
  (\bibinfo {year} {1953})}\BibitemShut {NoStop}%
\bibitem [{\citenamefont {Cartwright}\ and\ \citenamefont
  {Harary}(1956)}]{cartwright1956structural}%
  \BibitemOpen
  \bibfield  {author} {\bibinfo {author} {\bibfnamefont {D.}~\bibnamefont
  {Cartwright}}\ and\ \bibinfo {author} {\bibfnamefont {F.}~\bibnamefont
  {Harary}},\ }\href@noop {} {\bibfield  {journal} {\bibinfo  {journal}
  {Psychol. Rev.}\ }\textbf {\bibinfo {volume} {63}},\ \bibinfo {pages} {277}
  (\bibinfo {year} {1956})}\BibitemShut {NoStop}%
\bibitem [{\citenamefont {Marvel}\ \emph {et~al.}(2011)\citenamefont {Marvel},
  \citenamefont {Kleinberg}, \citenamefont {Kleinberg},\ and\ \citenamefont
  {Strogatz}}]{marvel2011continuous}%
  \BibitemOpen
  \bibfield  {author} {\bibinfo {author} {\bibfnamefont {S.~A.}\ \bibnamefont
  {Marvel}}, \bibinfo {author} {\bibfnamefont {J.}~\bibnamefont {Kleinberg}},
  \bibinfo {author} {\bibfnamefont {R.~D.}\ \bibnamefont {Kleinberg}}, \ and\
  \bibinfo {author} {\bibfnamefont {S.~H.}\ \bibnamefont {Strogatz}},\
  }\href@noop {} {\bibfield  {journal} {\bibinfo  {journal} {Proc. Natl. Acad.
  Sci. USA}\ }\textbf {\bibinfo {volume} {108}},\ \bibinfo {pages} {1771}
  (\bibinfo {year} {2011})}\BibitemShut {NoStop}%
\bibitem [{\citenamefont {Easley}\ and\ \citenamefont
  {Kleinberg}(2010)}]{easley2010networks}%
  \BibitemOpen
  \bibfield  {author} {\bibinfo {author} {\bibfnamefont {D.}~\bibnamefont
  {Easley}}\ and\ \bibinfo {author} {\bibfnamefont {J.}~\bibnamefont
  {Kleinberg}},\ }\enquote {\bibinfo {title} {A weaker form of structural
  balance},}\ in\ \href@noop {} {\emph {\bibinfo {booktitle} {Networks, Crowds,
  and Markets: Reasoning about a Highly Connected World}}}\ (\bibinfo
  {publisher} {Cambridge University Press},\ \bibinfo {address} {Cambridge},\
  \bibinfo {year} {2010})\ Chap.~\bibinfo {chapter} {5}, pp.\ \bibinfo {pages}
  {115--118}\BibitemShut {NoStop}%
\bibitem [{\citenamefont {Szell}\ \emph {et~al.}(2010)\citenamefont {Szell},
  \citenamefont {Lambiotte},\ and\ \citenamefont
  {Thurner}}]{szell2010multirelational}%
  \BibitemOpen
  \bibfield  {author} {\bibinfo {author} {\bibfnamefont {M.}~\bibnamefont
  {Szell}}, \bibinfo {author} {\bibfnamefont {R.}~\bibnamefont {Lambiotte}}, \
  and\ \bibinfo {author} {\bibfnamefont {S.}~\bibnamefont {Thurner}},\
  }\href@noop {} {\bibfield  {journal} {\bibinfo  {journal} {Proc. Natl. Acad.
  Sci. USA}\ }\textbf {\bibinfo {volume} {107}},\ \bibinfo {pages} {13636}
  (\bibinfo {year} {2010})}\BibitemShut {NoStop}%
\bibitem [{\citenamefont {Leskovec}\ \emph {et~al.}(2010)\citenamefont
  {Leskovec}, \citenamefont {Huttenlocher},\ and\ \citenamefont
  {Kleinberg}}]{leskovec2010signed}%
  \BibitemOpen
  \bibfield  {author} {\bibinfo {author} {\bibfnamefont {J.}~\bibnamefont
  {Leskovec}}, \bibinfo {author} {\bibfnamefont {D.}~\bibnamefont
  {Huttenlocher}}, \ and\ \bibinfo {author} {\bibfnamefont {J.}~\bibnamefont
  {Kleinberg}},\ }in\ \href@noop {} {\emph {\bibinfo {booktitle} {Proc. SIGCHI
  Conf. Hum. Factor Comput. Syst.}}}\ (\bibinfo  {publisher} {Association for
  Computing Machinery},\ \bibinfo {address} {New York, NY},\ \bibinfo {year}
  {2010})\ pp.\ \bibinfo {pages} {1361--1370}\BibitemShut {NoStop}%
\bibitem [{\citenamefont {Malarz}\ and\ \citenamefont
  {Ho{\l}yst}(2022)}]{malarz2022mean}%
  \BibitemOpen
  \bibfield  {author} {\bibinfo {author} {\bibfnamefont {K.}~\bibnamefont
  {Malarz}}\ and\ \bibinfo {author} {\bibfnamefont {J.~A.}\ \bibnamefont
  {Ho{\l}yst}},\ }\href@noop {} {\bibfield  {journal} {\bibinfo  {journal}
  {Phys. Rev. E}\ }\textbf {\bibinfo {volume} {106}},\ \bibinfo {pages}
  {064139} (\bibinfo {year} {2022})}\BibitemShut {NoStop}%
\bibitem [{\citenamefont {Wo{\l}oszyn}\ and\ \citenamefont
  {Malarz}(2022)}]{woloszyn2022thermal}%
  \BibitemOpen
  \bibfield  {author} {\bibinfo {author} {\bibfnamefont {M.}~\bibnamefont
  {Wo{\l}oszyn}}\ and\ \bibinfo {author} {\bibfnamefont {K.}~\bibnamefont
  {Malarz}},\ }\href@noop {} {\bibfield  {journal} {\bibinfo  {journal} {Phys.
  Rev. E}\ }\textbf {\bibinfo {volume} {105}},\ \bibinfo {pages} {024301}
  (\bibinfo {year} {2022})}\BibitemShut {NoStop}%
\bibitem [{\citenamefont {Malarz}\ and\ \citenamefont
  {Wo{\l}oszyn}(2023)}]{malarz2023thermal}%
  \BibitemOpen
  \bibfield  {author} {\bibinfo {author} {\bibfnamefont {K.}~\bibnamefont
  {Malarz}}\ and\ \bibinfo {author} {\bibfnamefont {M.}~\bibnamefont
  {Wo{\l}oszyn}},\ }\href@noop {} {\bibfield  {journal} {\bibinfo  {journal}
  {Chaos}\ }\textbf {\bibinfo {volume} {33}},\ \bibinfo {pages} {073115}
  (\bibinfo {year} {2023})}\BibitemShut {NoStop}%
\bibitem [{\citenamefont {Ku{\l}akowski}\ \emph {et~al.}(2005)\citenamefont
  {Ku{\l}akowski}, \citenamefont {Gawro{\'n}ski},\ and\ \citenamefont
  {Gronek}}]{kulakowski2005heider}%
  \BibitemOpen
  \bibfield  {author} {\bibinfo {author} {\bibfnamefont {K.}~\bibnamefont
  {Ku{\l}akowski}}, \bibinfo {author} {\bibfnamefont {P.}~\bibnamefont
  {Gawro{\'n}ski}}, \ and\ \bibinfo {author} {\bibfnamefont {P.}~\bibnamefont
  {Gronek}},\ }\href@noop {} {\bibfield  {journal} {\bibinfo  {journal} {Int.
  J. Mod. Phys. C}\ }\textbf {\bibinfo {volume} {16}},\ \bibinfo {pages} {707}
  (\bibinfo {year} {2005})}\BibitemShut {NoStop}%
\bibitem [{\citenamefont {Fujimoto}\ and\ \citenamefont
  {Ohtsuki}(2024)}]{fujimoto2024leader}%
  \BibitemOpen
  \bibfield  {author} {\bibinfo {author} {\bibfnamefont {Y.}~\bibnamefont
  {Fujimoto}}\ and\ \bibinfo {author} {\bibfnamefont {H.}~\bibnamefont
  {Ohtsuki}},\ }\href@noop {} {\bibfield  {journal} {\bibinfo  {journal} {PRX
  Life}\ }\textbf {\bibinfo {volume} {2}},\ \bibinfo {pages} {023009} (\bibinfo
  {year} {2024})}\BibitemShut {NoStop}%
\bibitem [{\citenamefont {Ilany}\ \emph {et~al.}(2013)\citenamefont {Ilany},
  \citenamefont {Barocas}, \citenamefont {Koren}, \citenamefont {Kam},\ and\
  \citenamefont {Geffen}}]{ilany2013structural}%
  \BibitemOpen
  \bibfield  {author} {\bibinfo {author} {\bibfnamefont {A.}~\bibnamefont
  {Ilany}}, \bibinfo {author} {\bibfnamefont {A.}~\bibnamefont {Barocas}},
  \bibinfo {author} {\bibfnamefont {L.}~\bibnamefont {Koren}}, \bibinfo
  {author} {\bibfnamefont {M.}~\bibnamefont {Kam}}, \ and\ \bibinfo {author}
  {\bibfnamefont {E.}~\bibnamefont {Geffen}},\ }\href@noop {} {\bibfield
  {journal} {\bibinfo  {journal} {Anim. Behav.}\ }\textbf {\bibinfo {volume}
  {85}},\ \bibinfo {pages} {1397} (\bibinfo {year} {2013})}\BibitemShut
  {NoStop}%
\bibitem [{\citenamefont {Ilany}\ and\ \citenamefont
  {Akcay}(2016)}]{ilany2016social}%
  \BibitemOpen
  \bibfield  {author} {\bibinfo {author} {\bibfnamefont {A.}~\bibnamefont
  {Ilany}}\ and\ \bibinfo {author} {\bibfnamefont {E.}~\bibnamefont {Akcay}},\
  }\href@noop {} {\bibfield  {journal} {\bibinfo  {journal} {Nat. Commun.}\
  }\textbf {\bibinfo {volume} {7}},\ \bibinfo {pages} {12084} (\bibinfo {year}
  {2016})}\BibitemShut {NoStop}%
\bibitem [{\citenamefont {Brandt}\ \emph {et~al.}(2007)\citenamefont {Brandt},
  \citenamefont {Ohtsuki}, \citenamefont {Iwasa},\ and\ \citenamefont
  {Sigmund}}]{brandt2007survey}%
  \BibitemOpen
  \bibfield  {author} {\bibinfo {author} {\bibfnamefont {H.}~\bibnamefont
  {Brandt}}, \bibinfo {author} {\bibfnamefont {H.}~\bibnamefont {Ohtsuki}},
  \bibinfo {author} {\bibfnamefont {Y.}~\bibnamefont {Iwasa}}, \ and\ \bibinfo
  {author} {\bibfnamefont {K.}~\bibnamefont {Sigmund}},\ }in\ \href@noop {}
  {\emph {\bibinfo {booktitle} {Mathematics for Ecology and Environmental
  Sciences}}},\ \bibinfo {editor} {edited by\ \bibinfo {editor} {\bibfnamefont
  {Y.}~\bibnamefont {Takeuchi}}, \bibinfo {editor} {\bibfnamefont
  {Y.}~\bibnamefont {Iwasa}}, \ and\ \bibinfo {editor} {\bibfnamefont
  {K.}~\bibnamefont {Sato}}}\ (\bibinfo  {publisher} {Springer},\ \bibinfo
  {year} {2007})\ Chap.~\bibinfo {chapter} {3}, pp.\ \bibinfo {pages}
  {21--49}\BibitemShut {NoStop}%
\end{thebibliography}
%

\newpage
\onecolumngrid

\setcounter{page}{1}

\setcounter{secnumdepth}{2}
\appendix

\renewcommand{\thefigure}{A\arabic{figure}}
\setcounter{figure}{0}

\section{\skb{Four social norms related to balance theory among the leading eight}}
\label{app:leading}

\begin{table}
\caption{\skb{Characterization of the leading eight~\cite{brandt2007survey}.
An observer observes an interaction between a donor and a recipient, where the donor may choose between cooperation (C) and defection (D).
The observer assesses the donor in the following way: The observer's updated assessment $\alpha_{uXv}$ is either good (G) or bad (B), depending on the observer's existing assessment of the donor ($u \in \{\text{G},\text{B}\}$), the donor's behavior to the recipient ($X \in \{\text{C},\text{D}\}$), and the observer's assessment of the recipient ($v \in \{\text{G},\text{B}\}$).
}}
\begin{ruledtabular}
\begin{tabular}{c|cccccccc|cccc}
 & $\alpha_\text{GCG}$ & $\alpha_\text{GDG}$ & $\alpha_\text{GCB}$ & $\alpha_\text{GDB}$ &
 $\alpha_\text{BCG}$ & $\alpha_\text{BDG}$ & $\alpha_\text{BCB}$ & $\alpha_\text{BDB}$ &
 $\beta_\text{GG}$ & $\beta_\text{GB}$ & $\beta_\text{BG}$ & $\beta_\text{BB}$\\\hline
L1 & G & B & G & G & G & B & G & B & C & D & C & C\\
L2 (Consistent Standing) & G & B & B & G & G & B & G & B & C & D & C & C\\
L3 (Simple Standing) & G & B & G & G & G & B & G & G & C & D & C & D\\
L4 & G & B & G & G & G & B & B & G & C & D & C & D\\
L5 & G & B & B & G & G & B & G & G & C & D & C & D\\
L6 (Stern Judging) & G & B & B & G & G & B & B & G & C & D & C & D\\
L7 (Staying) & G & B & G & G & G & B & B & B & C & D & C & D\\
L8 (Judging) & G & B & B & G & G & B & B & B & C & D & C & D
\end{tabular}
\label{tab:eight}
\end{ruledtabular}
\end{table}

\skb{
Table~\ref{tab:eight} shows the complete list of the leading eight~\cite{ohtsuki2004should}. Among them, the four norms that have been studied in this work and in Ref.~\onlinecite{bae2024exact} are related in the following way.
\begin{equation}
\begin{tikzcd}[row sep=huge,column sep=width("bbbbbbbbbbbbbbbbb"),/tikz/ampersand replacement=\&]
\text{L6 (stern judging)} \arrow[r, "\alpha_\text{GCB}=\text{G}"]
\arrow[d, "\alpha_\text{BDB}=\text{B}"] \&  \text{L4} \arrow[d, "\alpha_\text{BDB}=\text{B}"] \\
\text{L8 (judging)} \arrow[r, "\alpha_\text{GCB}=\text{G}"] \& \text{L7 (staying)}
\end{tikzcd}
\end{equation}
Stern judging is the simplest norm among these four, described as
\begin{equation}
\sigma_{od}' = \sigma_{or} \cdot \sigma_{dr}.
\label{eq:stern}
\end{equation}
As explained in the main text, this rule means that a donor decides an action toward the recipient according to $\sigma_{dr}$, and that an observer judges the donor as good only if the donor's decision coincides with the observer's own opinion about the recipient $\sigma_{or}$.
By adding an exception of $\alpha_\text{BDB}=\text{B}$ here, we obtain the rule of L8 (judging) as in Eq.~\eqref{eq:judging}. Or, by adding an exception of $\alpha_\text{GCB}=\text{G}$ to Eq.~\eqref{eq:stern}, we obtain the rule of L4. Thus, if an L4 player sees a good donor helping a bad recipient, the player does not change his or her opinion about the donor \textemdash It could just be a sign of naivety rather than of evil. Finally, L7 (staying) is obtained by applying both of these exceptions to Eq.~\eqref{eq:stern}, and this is why we said in the main text that L7 is for L8 what L4 is for L6.}

\begin{figure}
    \centering
    \includegraphics[width=0.5\linewidth]{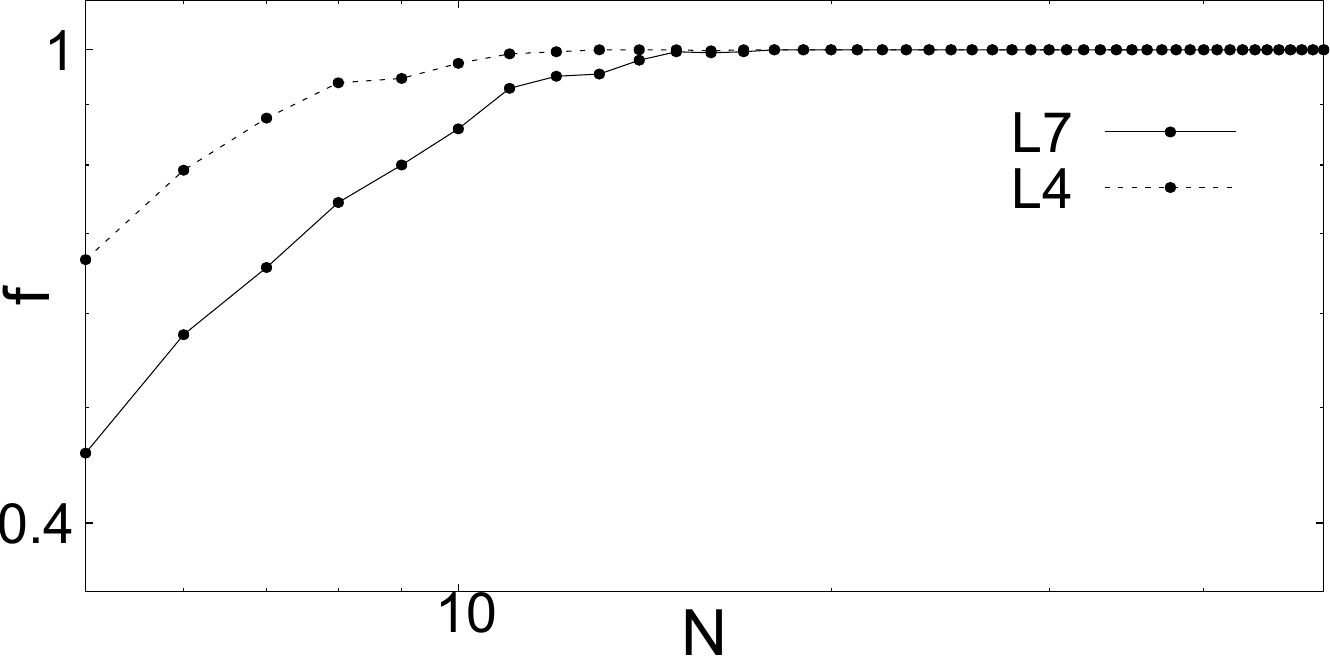}
    \caption{\skb{The fraction of samples reaching paradise, when started from independent random configurations $5\times10^2$ times, under the action of L7 or L4. As the total number of vertices $N$ increases, the fraction converges to $100\%$ in either case.}}
    \label{fig:paradise}
\end{figure}

\skb{Recall that L8 allows only individuals to move between clusters as a result of an assessment error. If we look at the cluster dynamics induced by L7, its difference from L8, i.e., $\alpha_\text{GCB}=\text{G}$, allows two clusters to merge due to a single assessment error. More specifically, consider the following cluster configuration:
\begin{equation}
\left\{ \left\{ 0, 1, 2\right\}, \left\{ 3, 4\right\}, \left\{ 5\right\} \right\}.
\label{eq:example}
\end{equation}
If node $0$ misjudges $3$ as good, the system under L8 will end up with one of the following three absorbing states: The first is the original configuration. The second is $\left\{ \left\{ 0\right\}, \left\{ 1, 2\right\}, \left\{ 3, 4\right\}, \left\{ 5\right\} \right\}$, which occurs with probability $P(3,2)$. The last is migration, which results in $\left\{ \left\{ 1, 2\right\}, \left\{ 0, 3, 4\right\}, \left\{ 5\right\} \right\}$ with probability $R(3,2)$. However, if L7 is the governing norm, it has one more possibility in addition to the above three, so the final configuration can be $\left\{ \left\{ 0, 1, 2, 3, 4\right\}, \left\{ 5\right\} \right\}$ as the two clusters merge. As shown in Fig.~\ref{fig:paradise}, this cluster-wise merging process makes the broad distribution as depicted in Fig.~\ref{fig:distribution}(b) collapse to paradise where all clusters have merged to one, as long as the system size $N$ is sufficiently greater than $O(10)$.
}

\renewcommand{\thefigure}{B\arabic{figure}}
\setcounter{figure}{0}

\section{Derivation of $P^\ast(m) = Q(m-1) = 1/m$}
\label{app:Pm}

\begin{figure}
\includegraphics[width=\textwidth]{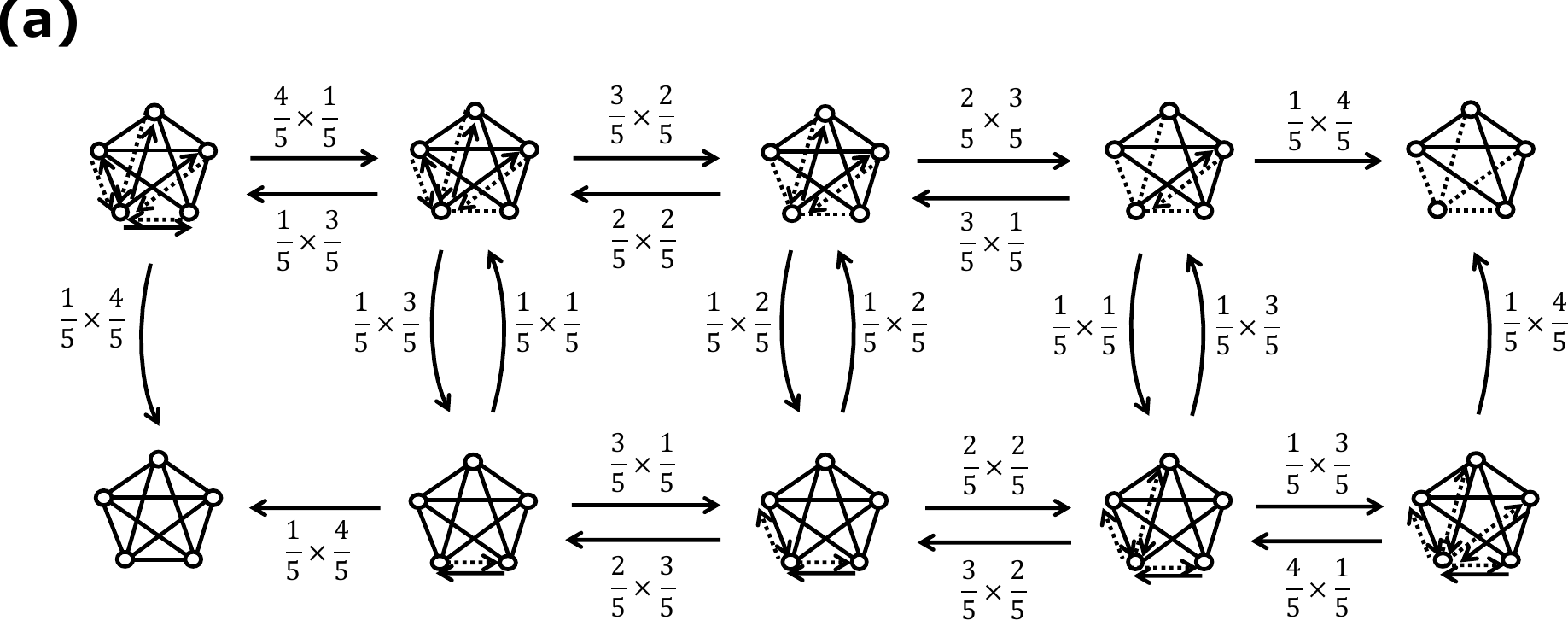}
\includegraphics[width=\textwidth]{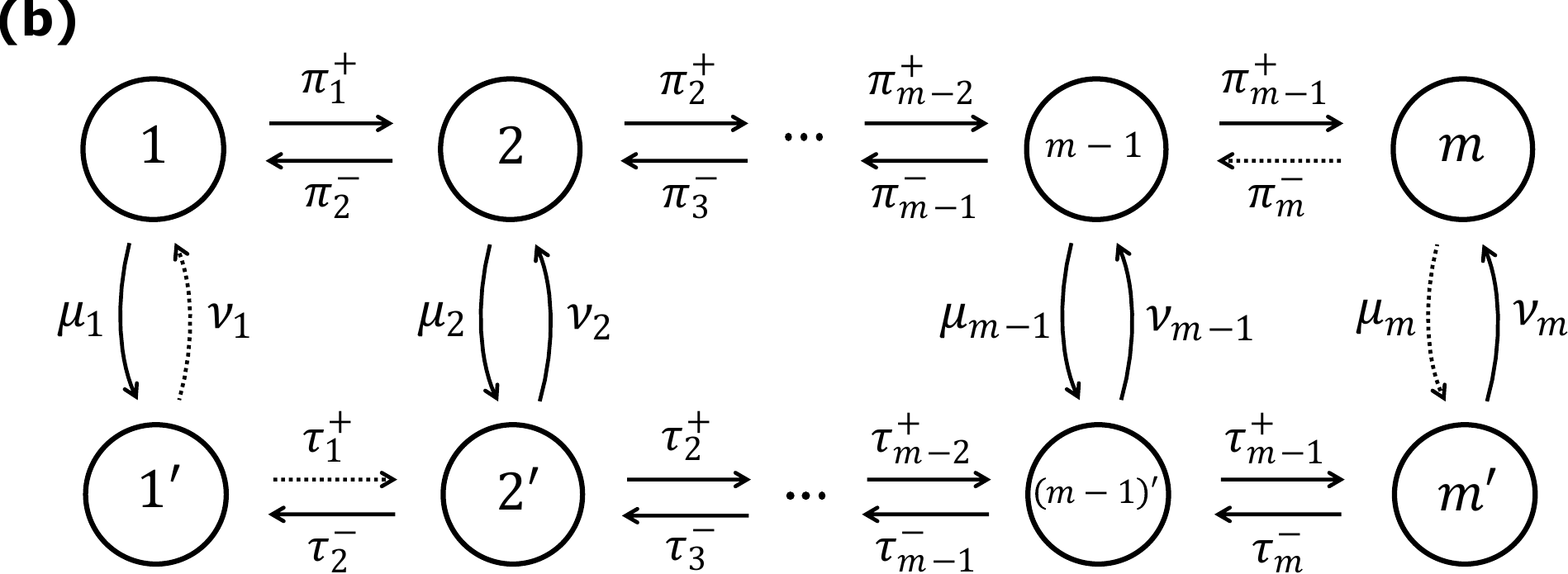}
\caption{Transition structure when a cluster of size $m$ is split into two because of an internal error between its member vertices. (a) An example of $m=5$, where each transition is represented by an arrow with its probability. In each configuration, solid and dotted arrows mean good and bad assessments, respectively. (b) Generalization to an arbitrary $m$. The transition probabilities are given in Eq.~\eqref{eq:PmTransition}, and we have drawn dotted arrows for $\nu_1$, $\tau_1^+$, $\pi_m^-$, and $\mu_m$ because the probabilities are actually zero.}
\label{fig:Pm}
\end{figure}

When an assessment error occurs in a cluster of size $m$, the transition between configurations forms a ladder structure [see Fig.~\ref{fig:Pm}(a) for $m=5$]. For general $m$, we have Fig.~\ref{fig:Pm}(b), where
\begin{subequations}
\begin{align}
\mu_j &\equiv \left(\frac{1}{m}\right) \left(\frac{m-j}{m}\right)\\
\pi_j^+ &\equiv \left(\frac{m-j}{m}\right) \left(\frac{j}{m}\right)\\
\pi_j^- &\equiv \left(\frac{j-1}{m}\right) \left(\frac{m-j}{m}\right)\\
\nu_j &\equiv \left(\frac{1}{m}\right) \left(\frac{j-1}{m}\right)\\
\tau_j^+ &\equiv \left(\frac{m-j}{m}\right) \left(\frac{j-1}{m}\right)\\
\tau_j^- &\equiv \left(\frac{j-1}{m}\right) \left(\frac{m-j+1}{m}\right).
\end{align}
\label{eq:PmTransition}
\end{subequations}
Each of observable configurations during the subsequent process is represented by a circle in Fig.~\ref{fig:Pm}(b), and the upper and lower circles are denoted by $j$ and $j'$, respectively, where $j=1,\ldots,m$. Starting from one of those configurations, the probability of absorption into a fully separated configuration (represented by the upper rightmost circle, $m$) is denoted by $q_j$ or $q_{j'}$ accordingly. The probabilities are related to each other by the following recursion formulas:
\begin{subequations}
\begin{align}
q_j &= \mu_j q_{j'} + \pi_j^+ q_{j+1} + \pi_j^- q_{j-1} + (1-\mu_j-\pi_j^+ - \pi_j^-) q_j\\
q_{j'} &= \nu_j q_j + \tau_j^+ q_{(j+1)'} + \tau_j^- q_{(j-1)'} + (1-\nu_j-\tau_j^+ - \tau_j^-) q_{j'}
\end{align}
\end{subequations}
with $q_{1'} \equiv 0$ and $q_m \equiv 1$.
It is straightforward to verify that the above equations are satisfied by the following solution:
\begin{subequations}
\begin{align}
q_j &= \frac{j}{m}\\
q_{j'} &= \frac{j-1}{m}.
\end{align}
\label{eq:qj}
\end{subequations}
The conditional probability $P^\ast(m)$ that a vertex separates from its cluster of size $m$ corresponds to $q_{2'} = 1/m$, whereas the merging probability is $Q(m-1) = 1-q_{m-1} = 1/m$. Note that $\alpha_{BDB}$, the only difference between L6 and L8, is not involved in this process at all, which means that $P^\ast(m) = Q(m-1)$ due to the path-reversal symmetry~\cite{bae2024exact}.

\renewcommand{\thefigure}{C\arabic{figure}}
\setcounter{figure}{0}

\section{Calculation of $P(m,n)$ and $R(m,n)$}
\label{app:Pmn}

Consider two clusters of respective sizes $m$ and $n$ with $m+n \le N$, where $N$ is the total number of vertices in the complete graph. If a member of the $m$-sized cluster, say $v_i$, makes an error in assessing a member of the $n$-sized cluster, we have three accessible absorbing configurations: The first is the original. The second is such that $v_i$ forms a new single-vertex cluster \skb{[Fig.~\ref{fig:eqs}(c)]}. The last is such that $v_i$ migrates to the $n$-sized cluster \skb{[Fig.~\ref{fig:eqs}(d)]}. The probability of absorption into each of these configurations is calculated in a numerically exact manner, as will be explained below.

\begin{figure}[ht]
\includegraphics[width=0.8\textwidth]{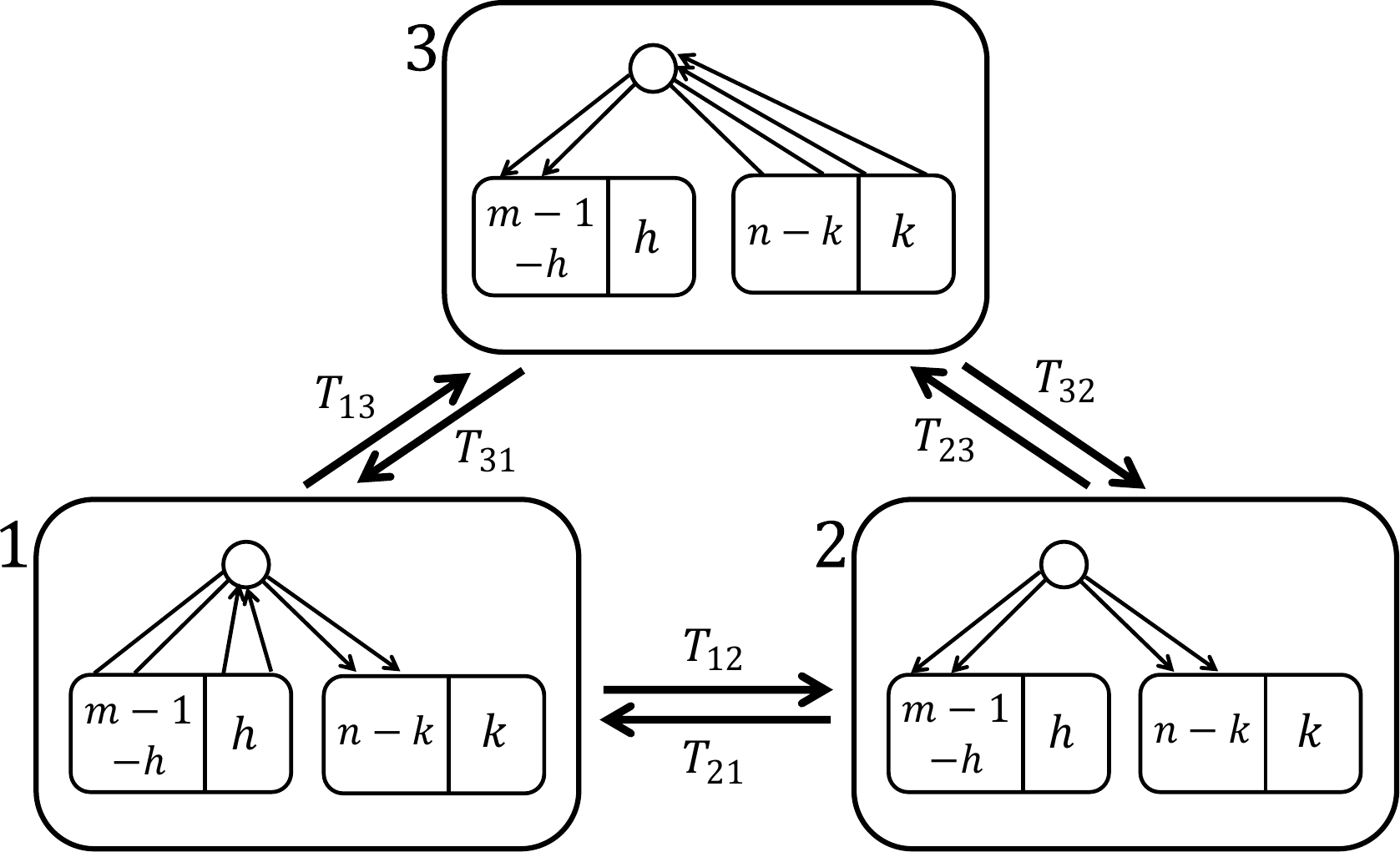}
\caption{A unit triangle for $0 \le k \le n-1$ and $0 \le h \le m-2$. The circle mean $v_i$ in $C'$ or $C''$, who was a member of the $m$-sized cluster in the original configuration but committed an assessment error toward a member of the $n$-sized cluster. We have drawn only positive links, and the links without arrow heads are bidirectional. The transition probabilities are given in Eq.~\eqref{eq:Tij}.}
\label{fig:cycle}
\end{figure}

\begin{figure}[ht]
\includegraphics[width=0.80\textwidth]{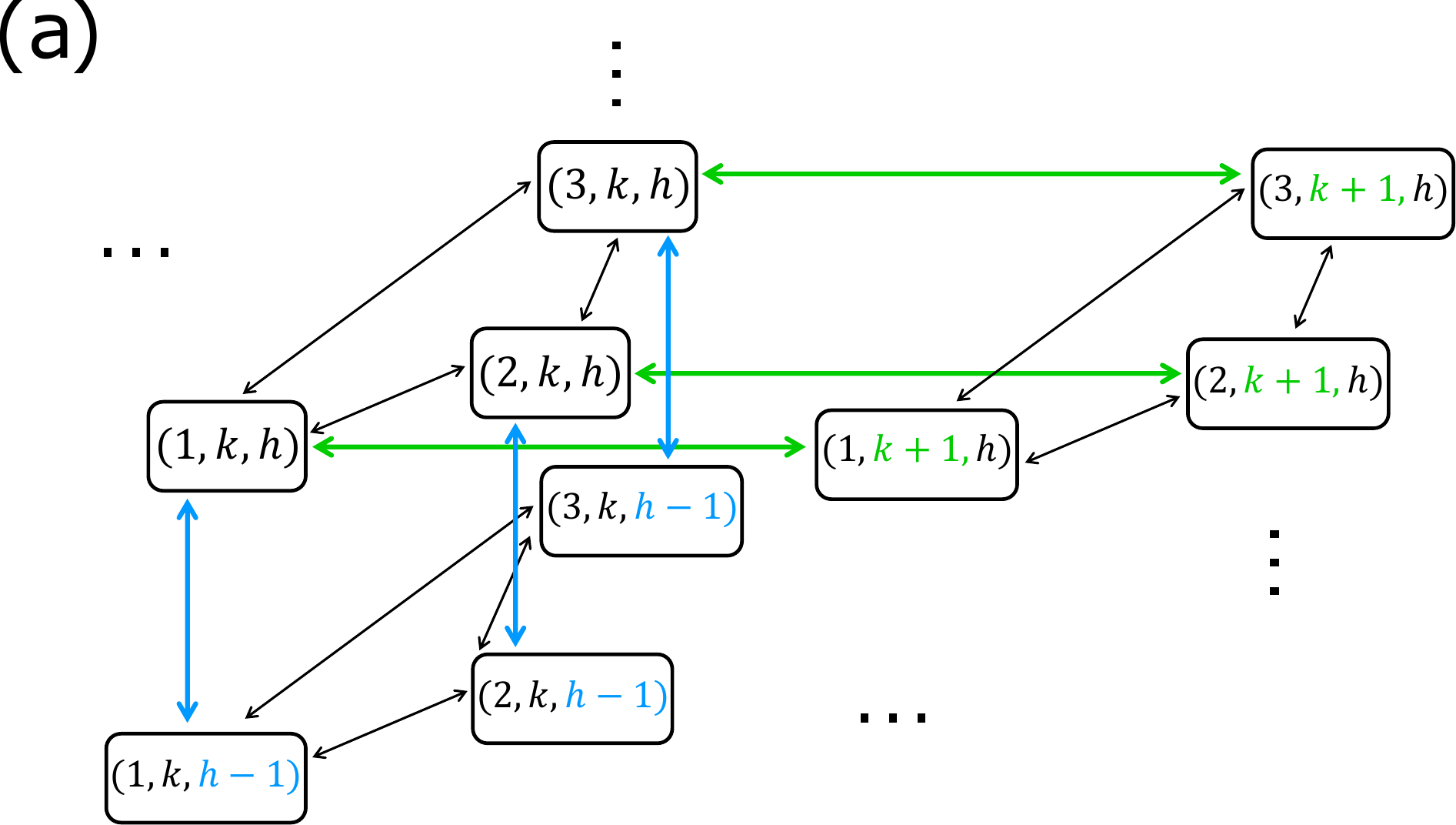}
\includegraphics[width=0.49\textwidth]{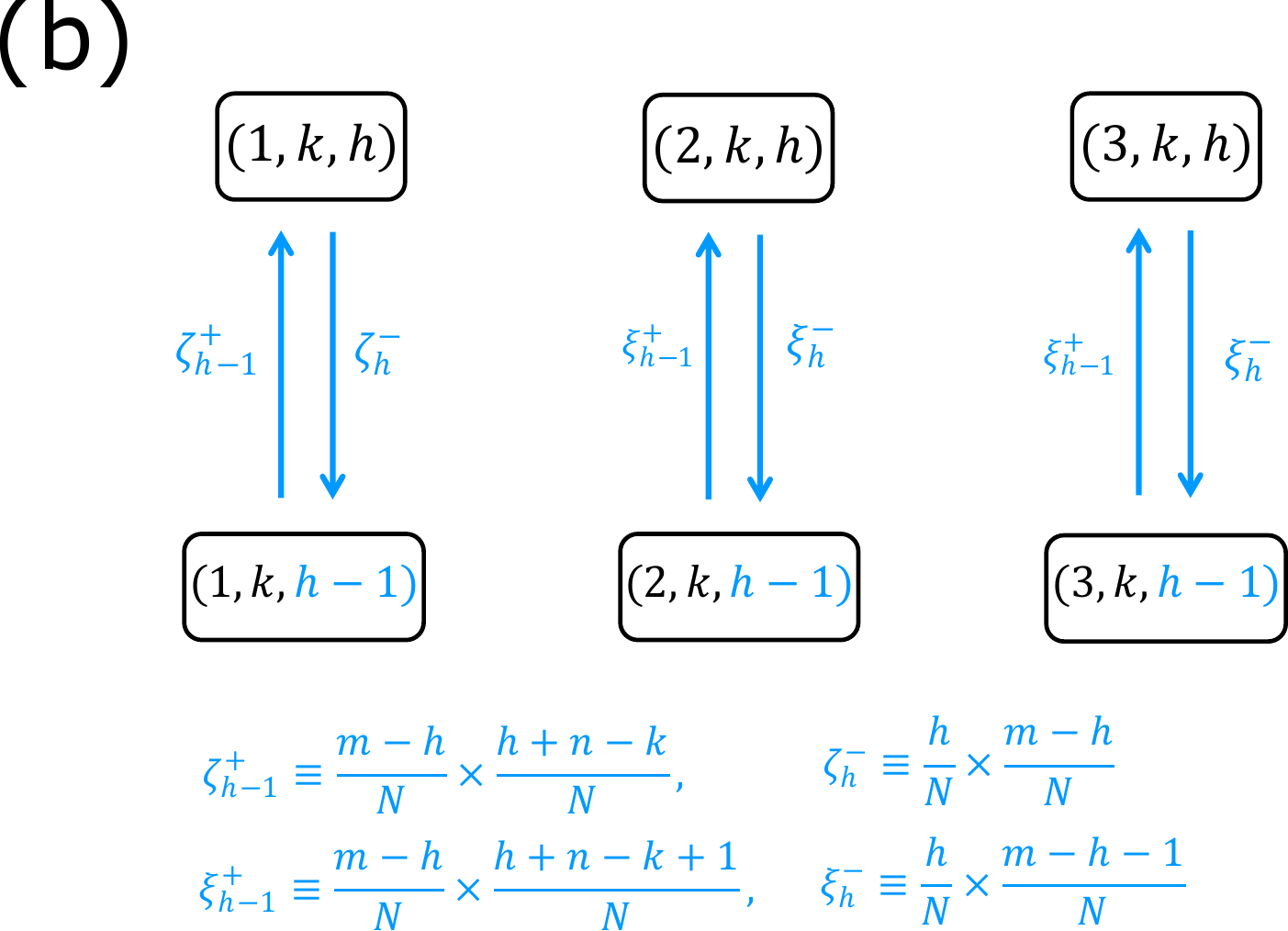}
\includegraphics[width=0.49\textwidth]{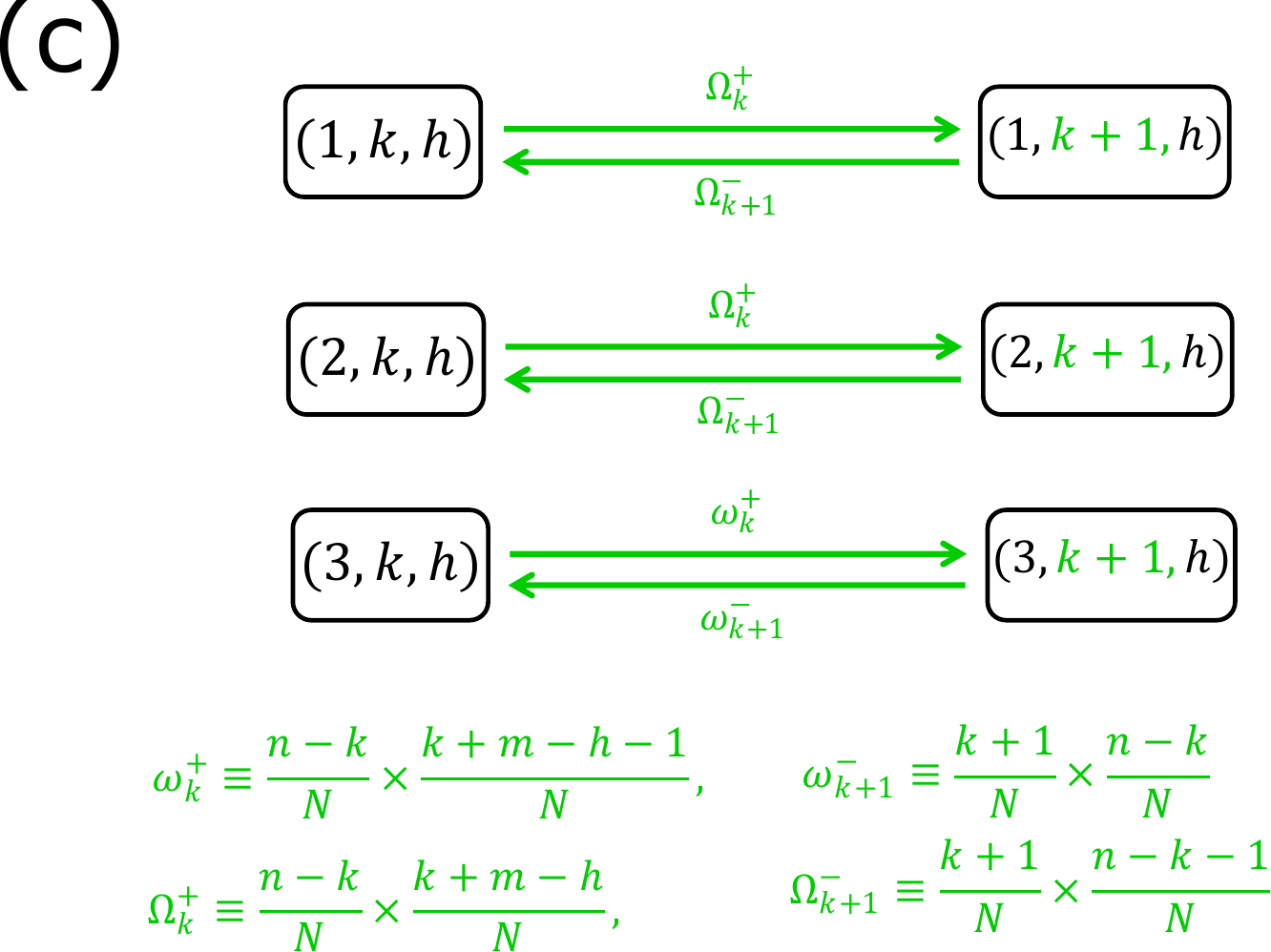}
\caption{(a) Three dimensional structure composed of triangular units such as in Fig.~\ref{fig:cycle}. (b) In the vertical direction, the unit triangle of $(1,k,h)$, $(2,k,h)$, and $(3,k,h)$ connects to other triangles having $h \pm 1$ with transition probabilities $\zeta^{\pm}_h$ or $\xi^{\pm}_h$. (c) In the horizontal direction, the unit triangle connects to other triangles having $k \pm 1$ with transition probabilities $\omega^{\pm}_k$ or $\Omega^{\pm}_k$.}
\label{fig:3Dstructure}
\end{figure}

\begin{figure}
\includegraphics[width=\textwidth]{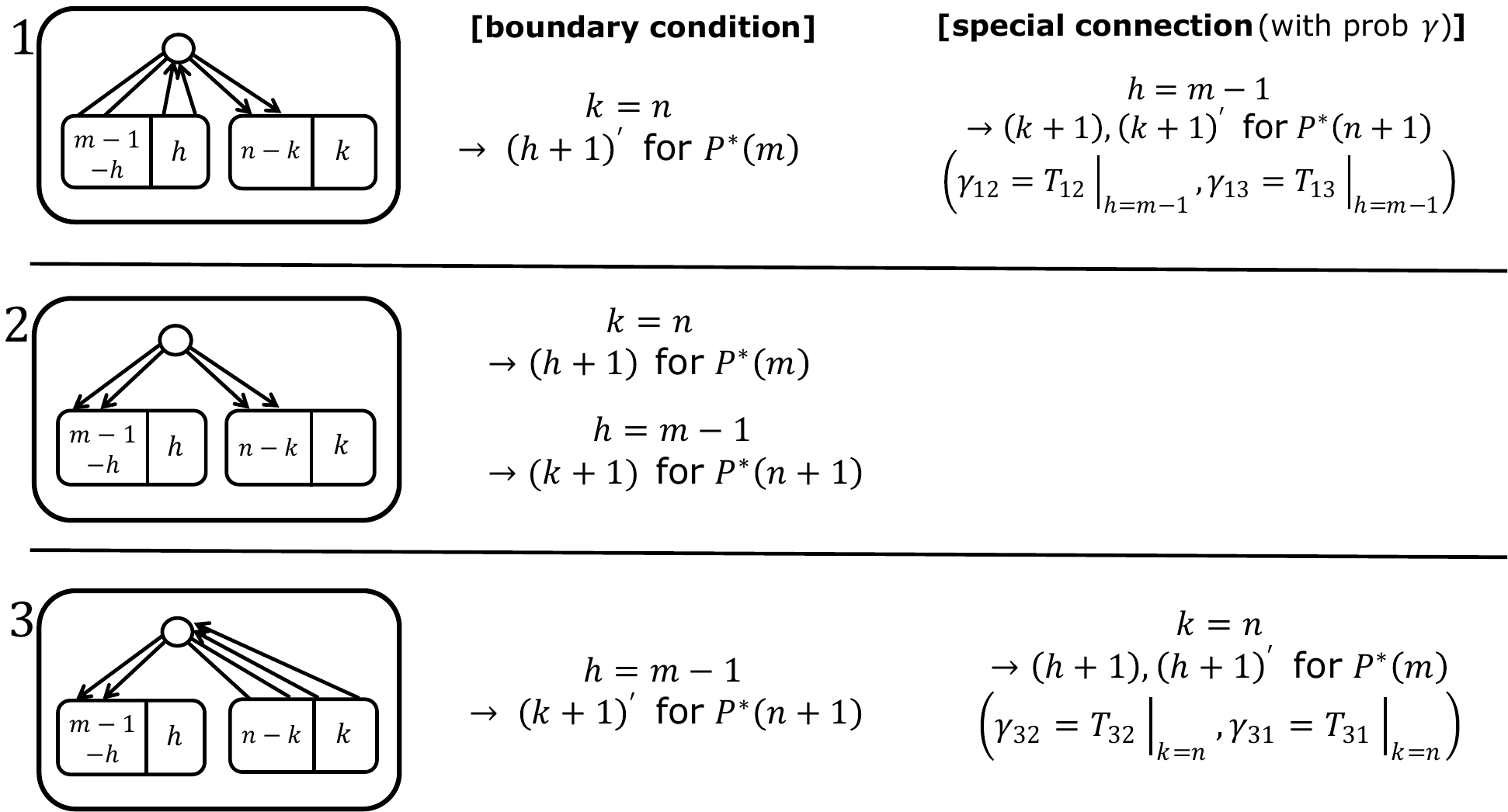}
\caption{Boundary conditions of the three-dimensional structure in Fig.~\ref{fig:3Dstructure}. As in Fig.~\ref{fig:cycle}, we have depicted only positive links, and those with arrow heads are bidirectional links.}
\label{fig:boundary}
\end{figure}

A crucial observation is the existence of a triangular unit consisting of the three configurations in Fig.~\ref{fig:cycle}, where the configurations can be visited with the following transition probabilities:
\begin{subequations}
\begin{align}
T_{12} &= \frac{1}{N} \times \frac{h}{N}\\
T_{21} &= \frac{1}{N} \times \frac{m-1-h}{N}\\
T_{23} &= \frac{1}{N} \times \frac{n-k}{N}\\
T_{32} &= \frac{1}{N} \times \frac{k}{N}\\
T_{31} &= \frac{1}{N} \times \frac{m-1-h}{N}\\
T_{13} &= \frac{1}{N} \times \frac{n-k}{N},
\end{align}
\label{eq:Tij}
\end{subequations}
where $0 \le k \le n-1$ and $0 \le h \le m-2$. This triangular unit can be characterized by two integers, $h$ and $k$. As shown in Fig.~\ref{fig:cycle}, the three configurations can thus be indicated by $(1,k,h)$, $(2,k,h)$, and $(3,k,h)$.
Then, such triangular units are connected to each other to form a three-dimensional structure as depicted in Fig.~\ref{fig:3Dstructure}(a).
In Fig.~\ref{fig:3Dstructure}(b) and (c), we have written the transition probabilities connecting the triangular units, denoted as $\zeta^\pm_h$, $\omega^\pm_k$, and $\Omega^\pm_k$.
Consequently, the absorption probabilities are related to each other by the following set of linear equations:
\begin{subequations}
\begin{align}
q_{1,k,h} &= T_{12} q_{2,k,h} + T_{13} q_{3,k,h} + \Omega^-_k q_{1,k-1,h} + \Omega^+_k q_{1,k+1,h} + \zeta^-_h q_{1,k,h-1} + \zeta^+_h q_{1,k,h+1}\nonumber\\
& + \left( 1-T_{12} - T_{13} - \Omega^-_k - \Omega^+_k - \zeta^-_h - \zeta^+_h \right) q_{1,k,h}\\
q_{2,k,h} &= T_{21} q_{1,k,h} + T_{23} q_{3,k,h} + \Omega^-_k q_{2,k-1,h} + \Omega^+_k q_{2,k+1,h} + \xi^-_h q_{2,k,h-1} + \xi^+_h q_{2,k,h+1}\nonumber\\
& + \left( 1-T_{21} - T_{23} - \Omega^-_k - \Omega^+_k - \xi^-_h - \xi^+_h \right) q_{2,k,h}\\
q_{3,k,h} &= T_{31} q_{1,k,h} + T_{32} q_{2,k,h} + \omega^-_k q_{3,k-1,h} + \omega^+_k q_{3,k+1,h} + \xi^-_h q_{3,k,h-1} + \xi^+_h q_{3,k,h+1}\nonumber\\
& + \left( 1-T_{31} - T_{32} - \omega^-_k - \omega^+_k - \xi^-_h - \xi^+_h \right) q_{3,k,h}.
\end{align}
\end{subequations}
We can obtain $P(m,n)$ and $Q(m,n)$ by solving this linear system.
Note that the three-dimensional structure is bounded by the ladder-shaped modules analyzed in Appendix~\ref{app:Pm}. One module is for a cluster of size $m$, and the other is for a cluster of size $(n+1)$. The absorption probability of each configuration inside the modules [Eq.~\eqref{eq:qj}] thus defines the boundary conditions of this three-dimensional random-walk problem with absorbing boundaries.
Let us decompose the boundary conditions into three parts.
The first is for $(1,k,h)$: if $k=n$, $(1,k,h)$ is mapped to a configuration that can be denoted by $(h+1)'$ in analyzing the cluster of size $m$. This is what we mean by ``$(h+1)'$ for $P^\ast(m)$'' in Fig.~\ref{fig:boundary}. In addition, if $h=m-1$, the system starting from $(1,k,h)$ can transit to $(k+1)$ for $P^\ast(n+1)$ with probability $\gamma_{12}= \left. T_{12} \right|_{h=m-1}$, or to $(k+1)'$ for $P^\ast(n+1)$ with probability $\gamma_{13}= \left. T_{13} \right|_{h=m-1}$.
The second part of the boundary conditions is for $(2,k,h)$: It corresponds to $(h+1)$ for $P^\ast(m)$ if $k=n$, and $(k+1)$ for $P^\ast(n+1)$ if $h=m-1$.
Finally, if $h=m-1$, $(3,k,h)$ corresponds to $(k+1)'$ for $P^\ast(n+1)$. In addition, if $k=n$, the system starting from $(3,k,h)$ can transit to $(h+1)$ for $P^\ast(m)$ with probability $\gamma_{32} = \left. T_{32} \right|_{k=n}$, or to $(h+1)'$ for $P^\ast(m)$ with probability $\gamma_{31} = \left. T_{31} \right|_{k=n}$.

\begin{figure}
    \centering
    \includegraphics[width=0.9\linewidth]{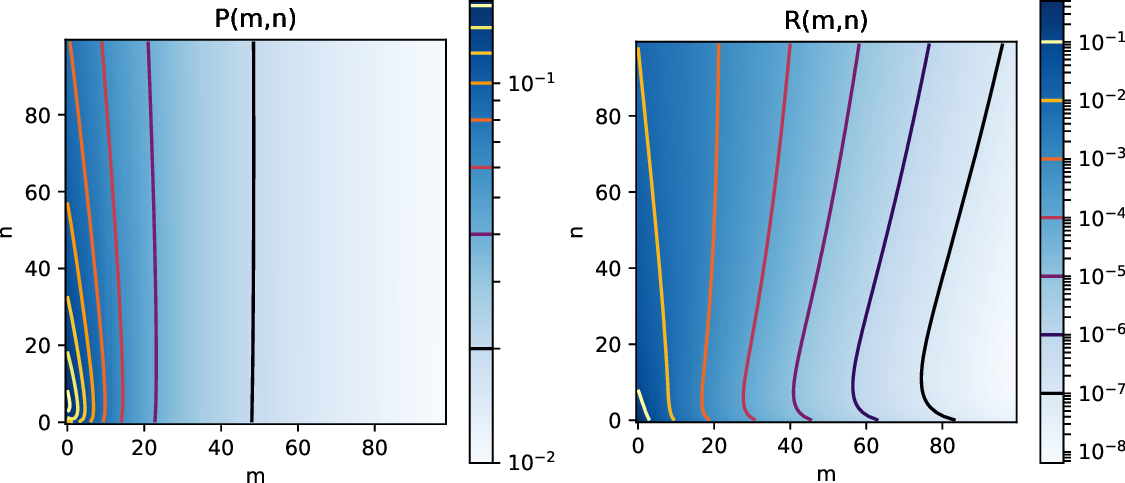}
    \caption{\skb{Contour plots of $P(m,n)$ and $R(m,n)$, obtained by calculating the absorption probabilities. We have included $R(1,n)=Q(n)=1/(n+1)$ in this plot.}}
    \label{fig:contour}
\end{figure}

\begin{figure}
\includegraphics[width=0.45\textwidth]{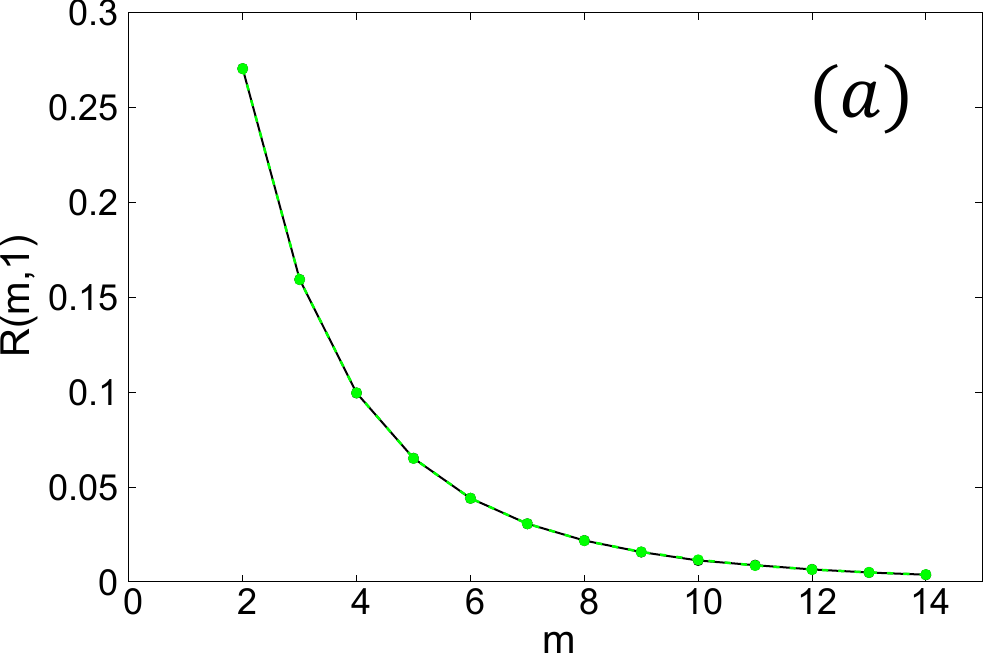}
\includegraphics[width=0.45\textwidth]{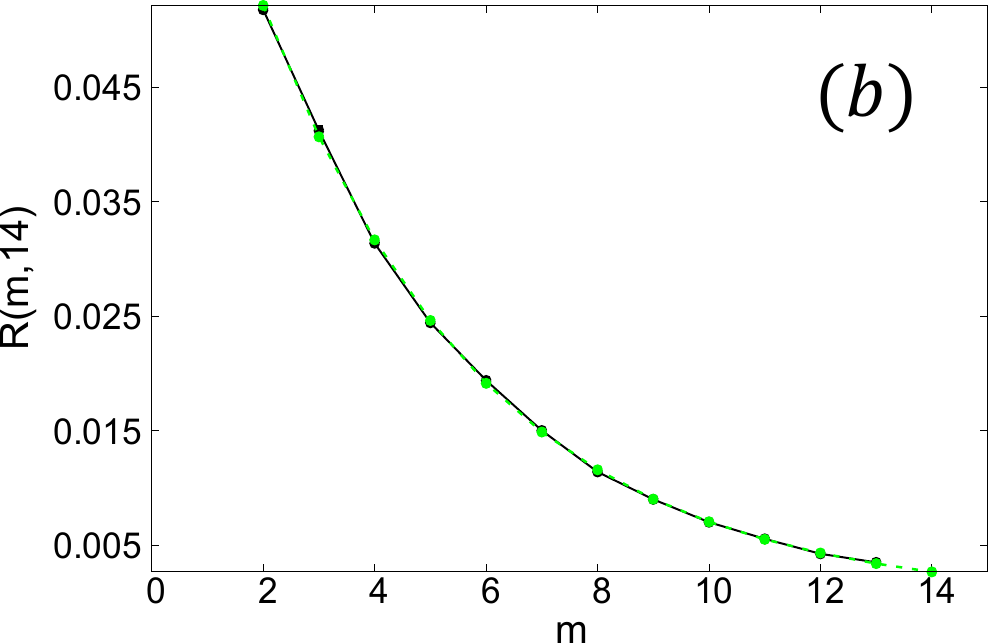}\\
\includegraphics[width=0.45\textwidth]{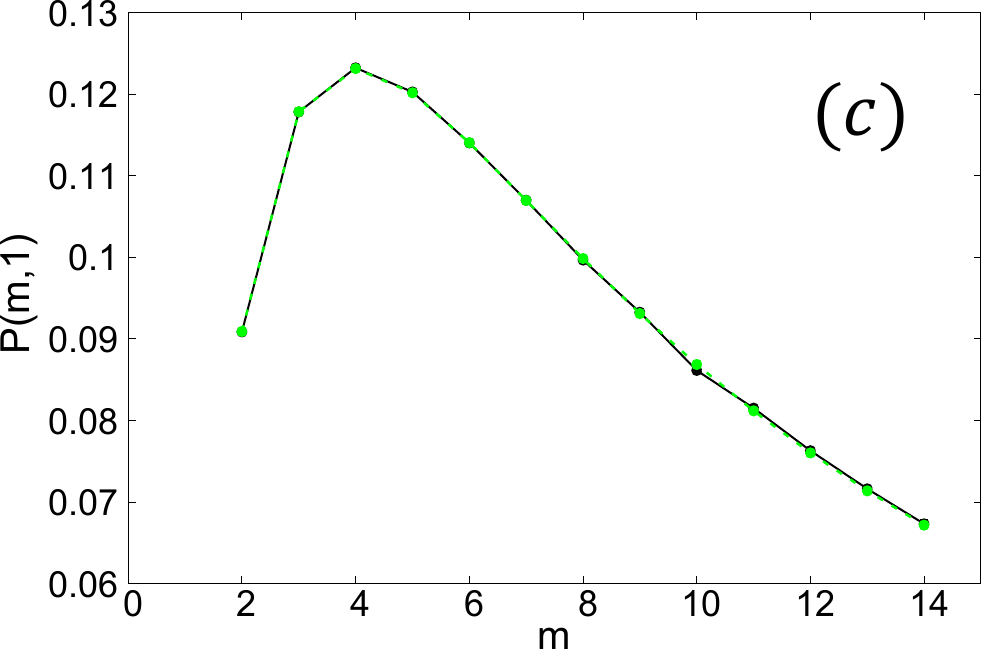}
\includegraphics[width=0.45\textwidth]{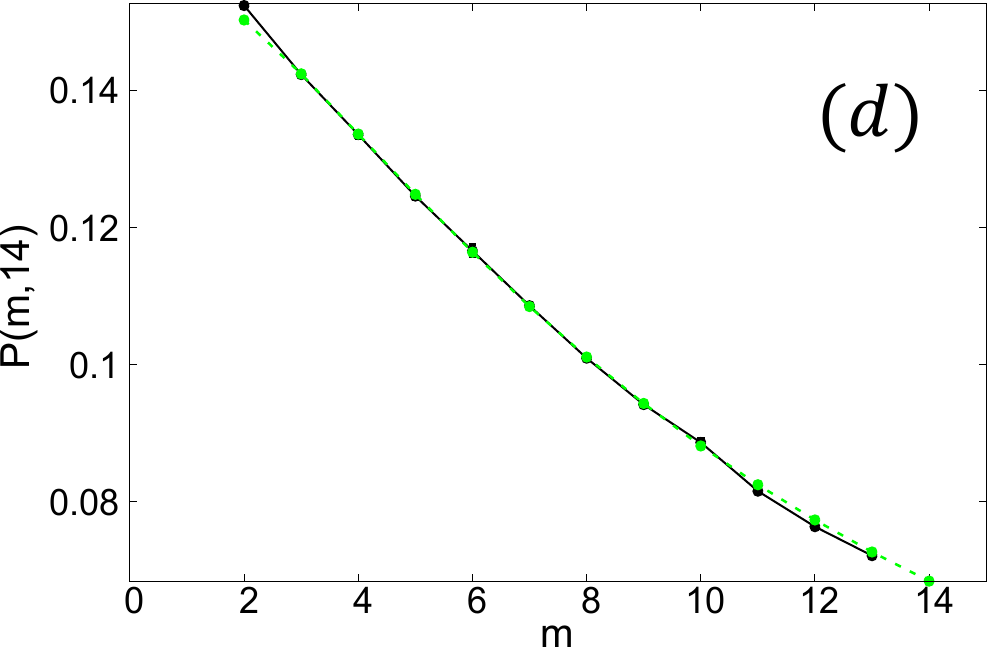}\\
\includegraphics[width=0.45\textwidth]{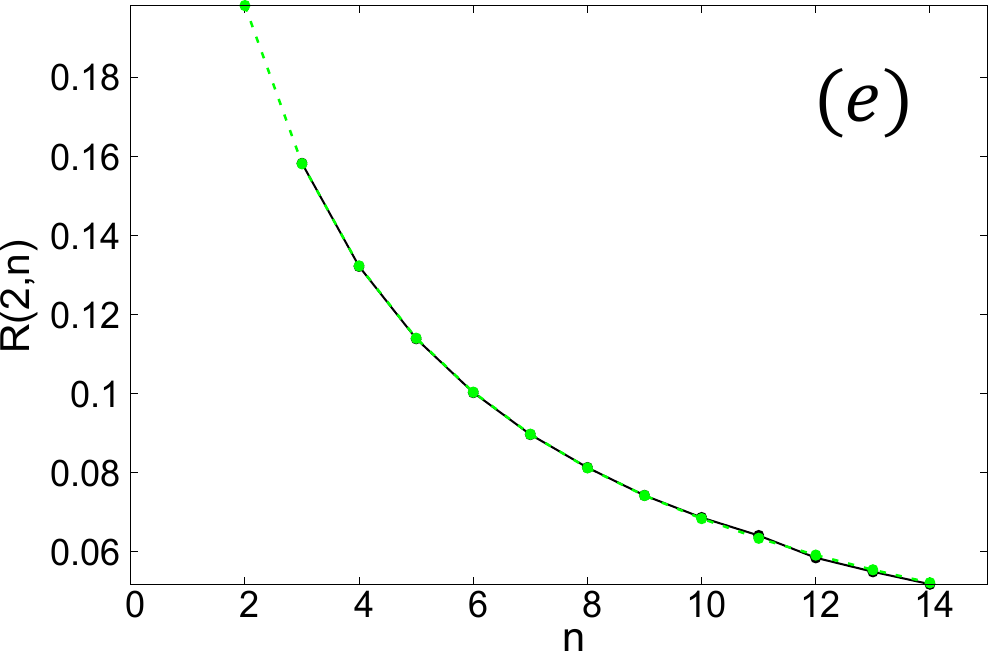}
\includegraphics[width=0.45\textwidth]{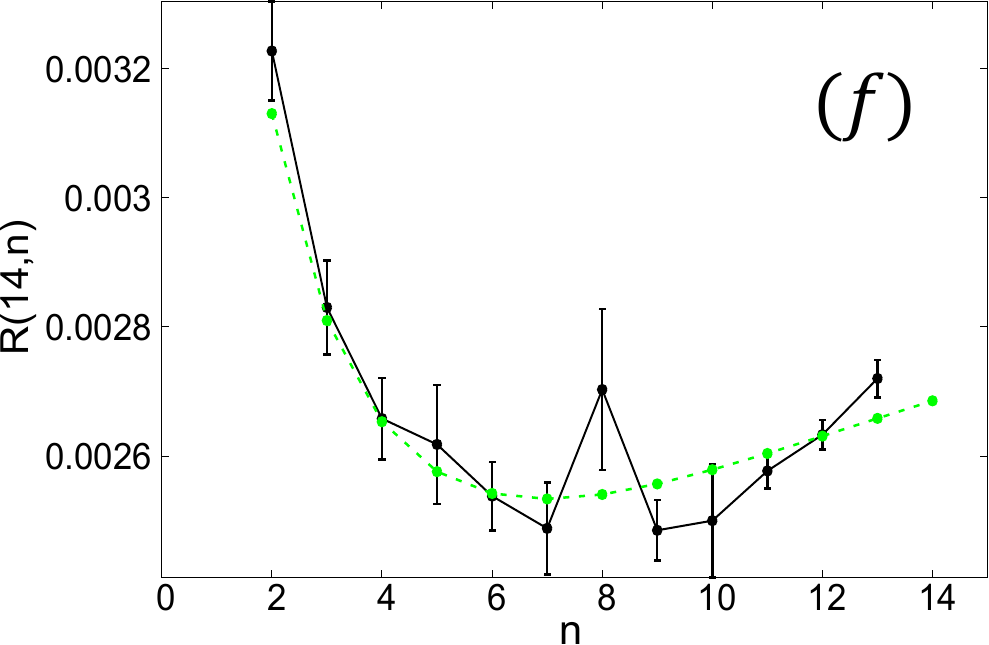}\\
\includegraphics[width=0.45\textwidth]{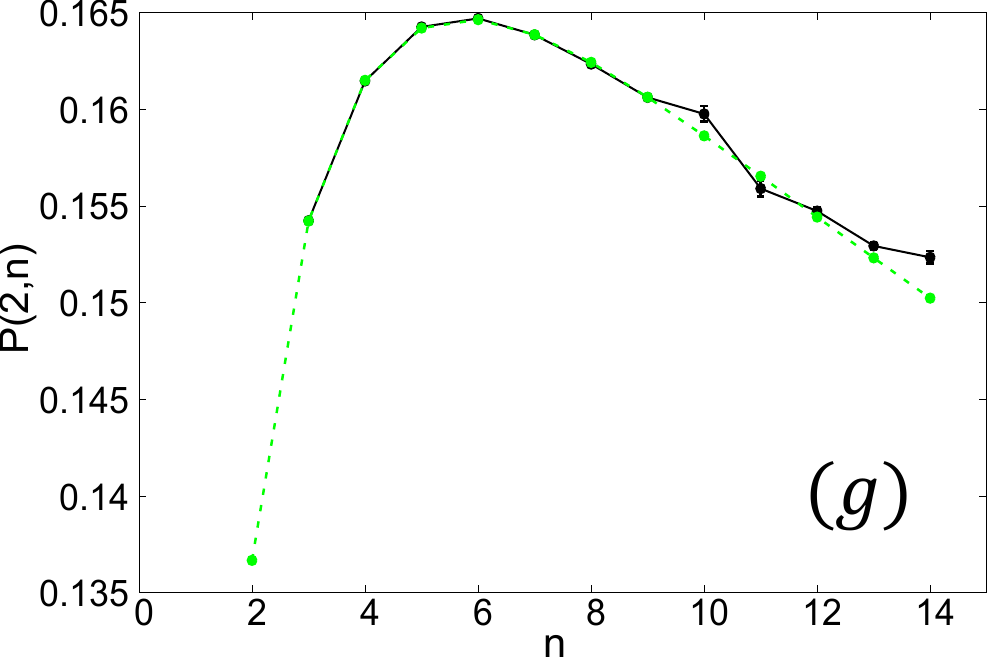}
\includegraphics[width=0.45\textwidth]{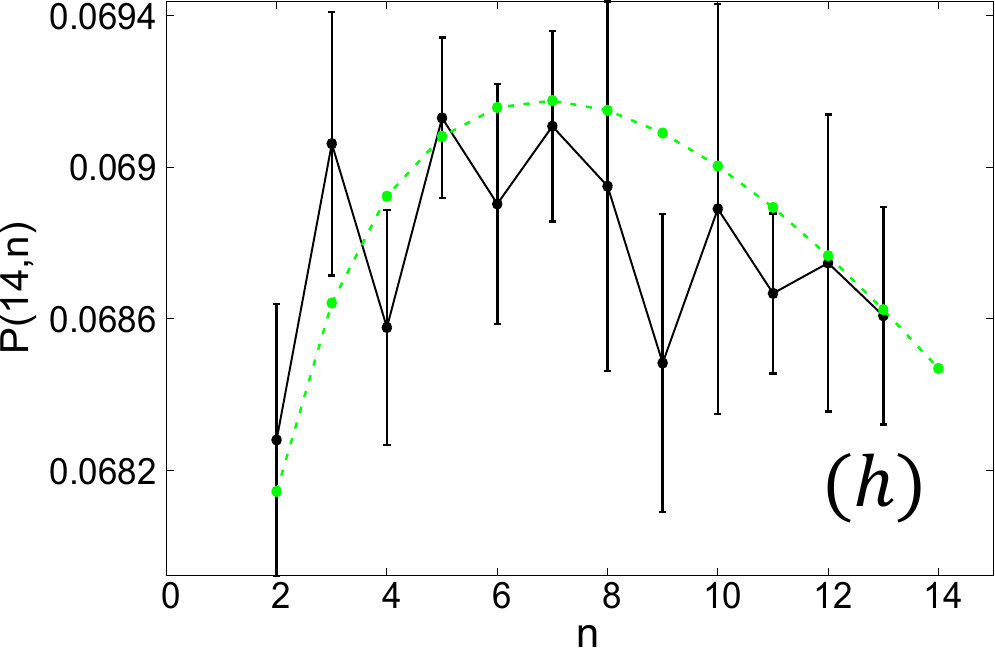}\\
\caption{Numerical confirmation of the transition probabilities calculated in Appendix~\ref{app:Pmn}. The black solid lines have been obtained from agent-based simulations according to the judging norm (Table~\ref{tab:eight}), and the green dotted lines show our numerically exact results.}
\label{fig:mc}
\end{figure}

\skb{Figure~\ref{fig:contour} shows the resulting probabilities on two-dimensional planes, and Fig.~\ref{fig:mc} compares the transition probabilities obtained in this way and Monte Carlo estimates from agent-based simulations.} Even when the size of a cluster is only $O(10)$, the transition probabilities are so small that the Monte Carlo estimates become highly imprecise [Fig.~\ref{fig:mc}(f) and (h)].

\renewcommand{\thefigure}{D\arabic{figure}}
\setcounter{figure}{0}

\section{Calculation of the size of the giant cluster}
\label{app:delta_nk}

In the main text, we have obtained Eq.~\eqref{eq:total} and another equation $\Delta n_1 = \Delta n_1^G + \Delta n_1^F - \Delta n_1^-=0$ [Eqs.~\eqref{eq:n1G} to \eqref{eq:n1-}] for analyzing the cluster dynamics. Now we consider finite clusters of size $k>1$. The number of finite clusters of size $k$ changes on average as follows:
\begin{eqnarray}
\Delta n_k &=& \sum_{k'=1,k'\neq k} \frac{(k'n_{k'}) [(k-1) n_{k-1}]}{N^2} R(k',k-1) \left( 1+\delta_{k',k+1}\right)\nonumber\\
&-& \sum_{k'=1,k'\neq k+1} \frac{(k'n_{k'}) (k n_k)}{N^2} R(k',k) \left( 1+\delta_{k',k}\right)\nonumber\\
&+&\sum_{k'=1,k'\neq k} \frac{[(k+1) n_{k+1}](k'n_{k'})}{N^2} R(k+1,k') \left( 1+\delta_{k,k'+1}\right)\nonumber\\
&-& \sum_{k'=1,k'\neq k-1} \frac{(kn_k) (k' n_{k'})}{N^2} R(k,k') \left( 1+\delta_{k',k}\right)\nonumber\\
&+& \frac{K (k-1) n_{k-1}}{N^2} R(K,k-1)- \frac{K k n_k}{N^2} R(K,k)\nonumber\\
&+& \frac{(k+1) n_{k+1}K}{N^2} R(k+1,K) - \frac{k n_k K}{N^2} R(k,K)\nonumber\\
&+& n_{k+1} \frac{(k+1)^2}{N^2} P^\ast(k+1) -
n_k \frac{k^2}{N^2} P^\ast(k)\nonumber\\
&+& \sum_{k'=1} \frac{(k+1) n_{k+1} (k'n_{k'})}{N^2} P(k+1,k')- \sum_{k'=1} \frac{kn_k (k'n_{k'})}{N^2} P(k,k').
\label{eq:nk}
\end{eqnarray}
As an approximation, we set a certain $k_\text{max}$, above which $n_k$ is assumed to be negligibly small. Then, we solve $\Delta n_k = 0$, together with Eq.~\eqref{eq:total}, to obtain $K$ and $n_k$ for $k=1,\ldots,k_\text{max}$.
One problem is that Eq.~\eqref{eq:nk} involves interactions with the giant cluster, whose size has yet to be determined. To handle this problem, we begin the calculation choosing $K$ in the probabilities as the largest possible value, for example, by replacing $R(k,K)$ by $R(k,N-k)$. Having solved the resulting set of equations, we obtain $K$ and $n_k$ for $k=1,\ldots,k_\text{max}$.
We then check whether Eq.~\eqref{eq:total} is satisfied. If so, we substitute this new $K$ into the equations and repeat the above calculations. Otherwise, we take $K = N-\sum_{k=1}^{k_\text{max}} n_k$. This iteration procedure ends when the solution converges.
\begin{figure}
\includegraphics[width=0.49\textwidth]{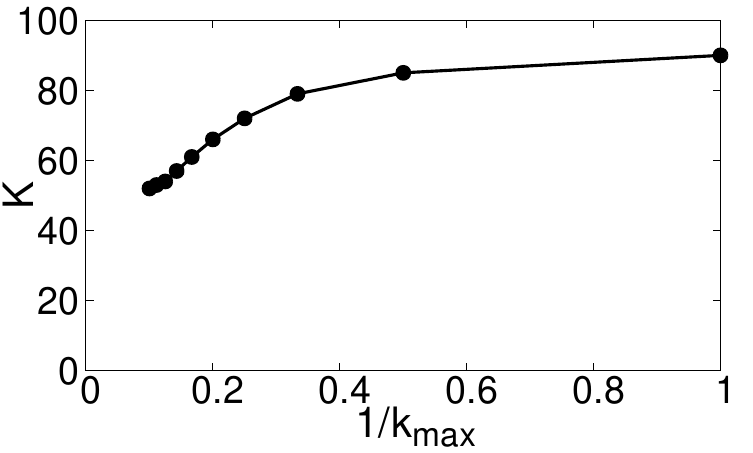}
\caption{The size of the giant cluster, $K$, plotted against $1/k_\text{max}$, where $k_\text{max}$ is the size of the largest finite clusters whose numbers are regarded as nonzero in our calculation. The total number of vertices is $N=10^2$.}
\label{fig:kmax}
\end{figure}
Figure~\ref{fig:kmax} shows how $K$ varies as $k_\text{max}$ increases when $N=10^2$. If we denote the characteristic scale of finite clusters by $k^\ast$, which is on the order of $10$ according to Fig.~\ref{fig:distribution}(a), the result will not change much once $k_\text{max}$ exceeds $k^\ast$.
Thus, our calculation is expected to converge to $K\approx 50$ as $k_\text{max}$ increases. This value is consistent with the observation in Fig.~\ref{fig:distribution}(a).

\end{document}